\documentclass[a4paper,11pt]{article}
\pdfoutput=1 

\usepackage{jcappub} 

\usepackage[T1]{fontenc} 

\usepackage{amsmath,esint}
\usepackage{yfonts}
\usepackage{units}
\usepackage{verbatim}
\usepackage{graphicx}
\usepackage{bm}
\usepackage{float}
\usepackage{caption}
\usepackage{subcaption}
\usepackage{hyperref}
\definecolor{mygray}{gray}{0.6}

\newcommand{\sv}{\langle \sigma \mathit{v} \rangle}
\newcommand{\svsvt}{\sv_\mathrm{svt}}
\newcommand{\svdsu}{\sv_{2.71}}

\newcommand{\data}{\bm{\mathcal{D}}}
\newcommand{\lkl}{\mathcal{L}}

\newcommand{\mdm}{m_\mathrm{DM}}

\newcommand{\dom}{{\Delta\Omega}}
\newcommand{\aeff}{ A_\mathrm{eff}}

\newcommand{\Jobs}{J_\mathrm{obs}}

\newcommand{\lp}{\lambda_P}

\newcommand{\Tobsi}{T_{\mathrm{obs},i}}

\newcommand{\Nonij}{{N_{\mathrm{ON},ij}}}
\newcommand{\Noffij}{{N_{\mathrm{OFF},ij}}}
\newcommand{\Nbins}{N_\mathrm{bins}}
\newcommand{\gij}{g_{ij}}
\newcommand{\bij}{b_{ij}}
\newcommand{\tauobsi}{\tau_{\mathrm{obs},i}}
\newcommand{\sigmataui}{\sigma_{\tau,i}}
\newcommand{\Eminj}{{E'_{\mathrm{min},j}}}
\newcommand{\Emaxj}{{E'_{\mathrm{max},j}}}

\title{\boldmath Indirect dark matter searches in the dwarf satellite galaxy Ursa~Major~II with the MAGIC Telescopes}

\author {MAGIC Collaboration:}
\author [1] {M.~L.~Ahnen}
\author [2,20] {S.~Ansoldi}
\author [3] {L.~A.~Antonelli}
\author [4] {C.~Arcaro}
\author [5] {D.~Baack}
\author [6] {A.~Babi\'c}
\author [7] {B.~Banerjee}
\author [8] {P.~Bangale}
\author [8,9] {U.~Barres de Almeida}
% ...

\affiliation [1] {ETH Zurich, CH-8093 Zurich, Switzerland}
\affiliation [2] {Universit\`a di Udine, and INFN Trieste, I-33100 Udine, Italy}
\affiliation [3] {National Institute for Astrophysics (INAF), I-00136 Rome, Italy}
\affiliation [4] {Universit\`a di Padova and INFN, I-35131 Padova, Italy}
\affiliation [5] {Technische Universit\"at Dortmund, D-44221 Dortmund, Germany}
\affiliation [6] {Croatian MAGIC Consortium: University of Rijeka, 51000 Rijeka, University of Split - FESB, 21000 Split,  University of Zagreb - FER, 10000 Zagreb, University of Osijek, 31000 Osijek and Rudjer Boskovic Institute, 10000 Zagreb, Croatia.}
\affiliation [7] {Saha Institute of Nuclear Physics, HBNI, 1/AF Bidhannagar, Salt Lake, Sector-1, Kolkata 700064, India}
\affiliation [8] {Max-Planck-Institut f\"ur Physik, D-80805 M\"unchen, Germany}
\affiliation [9] {now at Centro Brasileiro de Pesquisas F\'isicas (CBPF), 22290-180 URCA, Rio de Janeiro (RJ), Brasil}
% ...
\affiliation [20] {Japanese MAGIC Consortium: ICRR, The University of Tokyo, 277-8582 Chiba, Japan; Department of Physics, Kyoto University, 606-8502 Kyoto, Japan; Tokai University, 259-1292 Kanagawa, Japan; The University of Tokushima, 770-8502 Tokushima, Japan}
% ...

\emailAdd{paola.giammaria@oa-roma.inaf.it, saverio.lombardi@oa-roma.inaf.it, jpalacio@ifae.es, monica.vazquez.acosta@cern.ch}

\abstract{The dwarf spheroidal galaxy Ursa Major II (UMaII) is believed to be one of the most dark-matter dominated systems among the Milky Way satellites and represents a suitable target for indirect dark matter (DM) searches. The MAGIC telescopes carried out a deep observation campaign on UMaII between 2014 and 2016, collecting almost one hundred hours of good-quality data. This campaign enlarges the pool of DM targets observed at very high energy (E~$\gtrsim$~50~GeV) in search for signatures of DM annihilation in the wide mass range between $\sim$100~GeV and $\sim$100~TeV. To this end, the data are analyzed with the full likelihood analysis, a method based on the exploitation of the spectral information of the recorded events for an optimal sensitivity to the explored DM models. We obtain constraints on the annihilation cross-section for different channels that are among the most robust and stringent achieved so far at the TeV mass scale from observations of dwarf satellite galaxies.}


\begin{document}
\maketitle
\flushbottom

\section{Introduction}
\label{S:1}

Compelling evidence for a large ($\sim$85\%), dark, non-baryonic and non-relativistic (i.e. ``cold'') component of the matter density of the Universe arises at all astrophysical scales~\cite{Einasto14}. We infer its existence from the observations of gravitational effects on galaxies~\cite{Rubin70, Rubin80}, galaxy clusters~\cite{Zwicky33, Squires96, Bradac06} and from the anisotropies of the Cosmic Microwave Background~\cite{Planck15}. Despite the intensive and multi-approach efforts over the past decades, the nature of dark matter (DM) is still unknown and represents a paramount open issue of modern fundamental Physics and Astrophysics~\cite{Feng10}. A particularly well-motivated and widely considered class of cold DM particle candidates is the so-called Weakly Interacting Massive Particle (WIMP~\cite{Steigman12}). WIMPs spontaneously arise in many Standard Model~(SM) extensions (most notably Supersymmetry~\cite{Jungman96}), have interaction cross-sections typical of the weak scale and a mass in the range between $\sim$10~GeV and tens of TeV, and naturally provide the observed relic density (a fact popularly known as the ``WIMP miracle''~\cite{Salati14}).

Among different experimental approaches aimed at shedding light on DM nature~\cite{Balducci15, Atlas15,Xenon12,Lux13,DarkSide15,Feng17}, indirect searches~\cite{Gaskins16} look for SM particles (i.e. photons, cosmic rays, and neutrinos) produced by DM annihilation or decay processes in DM over-dense astrophysical regions.
Due to their complementarity in terms of energy coverage and sensitivity, the spaceborne and ground-based gamma-ray instruments~--~such as the imaging atmospheric Cherenkov telescopes~\cite{IACT15} (IACTs)~--~operate in synergy in order to scan the wide allowed WIMPs mass range ($\sim$10~GeV--$\sim$100~TeV), searching for a flux of gamma rays traced back to DM sources. In this respect, ground-based observations at very high energy (VHE, E~$\gtrsim$~50~GeV) are of major relevance in order to access the complementary parameter space of heavier (and well-motivated) DM masses with respect to the ones probed by spaceborne instruments at high energy (HE, E~$>$~100~MeV), as recently shown in the first ever joint DM analysis between \emph{Fermi}-LAT and MAGIC~\cite{ahnen16}. Furthermore, the current status of experimental searches  seems to strengthen the motivation for WIMPs with masses at the TeV scale or above~\cite{Livio14}. This is the mass range where IACTs provide the best sensitivity among all gamma-ray instruments, making this class of detectors particularly suited for indirect DM searches in the WIMP scenario.

Indirect DM signatures are expected to be observable in different classes of astrophysical objects, such as the Galactic Center (GC) and Galactic Halo (GH) regions~\cite{Van Eldik15}, galaxy clusters~\cite{Cirelli12}, and dwarf spheroidal satellite galaxies (dSphs) of the Milky Way~\cite{Martinez09, Persic96}. Over the last decade, all these classes of targets have been observed at VHE by the current generation of IACTs ~\cite{Doro14}(H.E.S.S.~\cite{Hess16, Abramowski14}, MAGIC~\cite{aleksic11, aleksic14, Doro17}, and VERITAS~\cite{Veritas15, Veritas17}), so far with no hints of DM signals. Nevertheless, stringent constraints to DM particle models in the TeV mass range have been set from these observations~\cite{ahnen16,Hess16}.

The dSph satellites of the Milky Way are among the best-suited targets for indirect DM searches to be observed by gamma-ray instruments (detailed reviews can be found e.g. in~\cite{Battaglia13, Ullio16}). So far, about thirty among faint and ultra-faint dwarf satellites have been identified by past (e.g. SDSS~\cite{Fukugita96}) and current (e.g. DES~\cite{des16}, Pan-STARRS~\cite{Chambers16}) deep optical sky surveys, and their number is expected to increase in the next years thanks to the on-going and future optical surveys~\cite{Antoja15}. This circumstance is of utmost importance for indirect DM searches since new discovered objects may show outstanding features and completely change our current DM detection prospects or capabilities to constrain models.

Ursa Major II~\cite{Zucker06} (UMaII) is believed to be one of the most DM dominated ultra-faint dSphs, with a Mass--to--Light ratio M/L~$\sim$~4000$^{+3700}_{-2000}$~$M_\odot/L_\odot$~\cite{Wolf10}. It has  an absolute magnitude of $M_V\sim-3.8$, and a distance of $\sim$~30~kpc (at RA (J2000)~=~8$^h$51$^{\prime}$30.0$^{\prime\prime}$ and Dec (J2000)~=~+63$^\circ$07'48''). According to kinematic studies~\cite{Simon07}, the maximum containment angle of DM emission is $\theta_{max} \simeq 0.53^\circ$\footnote{In the spherically symmetric model of dSph, the DM density profile  is a function of the halo-centric radius. A scale radius is representative of the extension of the innermost DM density profile. According to the kinematic data of member stars  it sets  a limit  beyond of which  the density profile is steeply falling and even the expected DM emission.  Thus an obvious choice for a conservative truncation radius of DM annihilation emission is that of the outermost member star ($r_{max}$) used to estimate the velocity dispersion profile. Then here $\theta_{max}$ is the angle corresponding to $r_{max}$, i.e. the median estimated distance of the outermost member star from the center of the system.}. Therefore, UMaII is an extended source compared to the typical IACT point spread function (PSF~$\sim$~0.1$^\circ$). This required special care for its observation with MAGIC and for the subsequent data analysis.

The MAGIC telescopes are carrying on deep campaigns for indirect DM searches on several selected sky regions~\cite{Giammaria16}. The diversification of targets is the optimal observational strategy pursued by MAGIC with the aim of reducing the uncertainties and biases in the selection of targets for indirect DM searches. The ultimate goal is to enhance the chances of positive detection and, in case of no hints of DM signal, to achieve the most robust and stringest limits at the TeV DM mass scale by means of the combination of results coming from different target observations. In this respect, the observation of UMaII belongs to a deep multi-year observation program on dSphs. Thanks to it, MAGIC already provided remarkable limits on DM particle models in the TeV mass range with the deep survey of Segue~1 dSph~\cite{aleksic14}. In this paper we present the results achieved by means of optimal DM analysis methods of the data taken by MAGIC in a two-years campaign on UMaII.

The rest of the paper is organized as follows: Section~\ref{S:2} provides details on the expected gamma-ray flux from the DM halo of UMaII. Section~\ref{S:3} introduces the MAGIC telescopes and the UMaII observation campaign considered in this study. The standard MAGIC data reduction procedure and results are reported in Section~\ref{S:4}. Then, in Section~\ref{S:5}, the full likelihood analysis method used to analyze the data is described. Section~\ref{S:6} presents the main results achieved by this study, i.e. the upper limits on annihilation cross section for different considered annihilation channels. Finally, the summary and conclusions of this work are given in Section~\ref{S:7}.

\section{Expected gamma-ray flux from the DM halo of UMaII}
\label{S:2}

The gamma-ray flux ($d\Phi/dE$) produced by DM annihilation (or decay) arriving at Earth
from a given region of the sky ($\dom$) is proportional to the product of two terms. The first one is the so-called \emph{particle physics} factor:
\begin{equation}
\frac{d\Phi^\mathrm{PP}}{dE} = \frac{1}{4\pi} \frac{\langle \sigma_\mathrm{ann}\mathit{v} \rangle}{2\, {\mdm}^2}\sum_i{Br_i\frac{dN_\gamma}{dE}}~~~
\label{eq:particlephysics}
\end{equation}
This factor contains all the information relative to the specific DM particle model, $ \langle \sigma_\mathrm{ann}\mathit{v} \rangle$ is the thermal averaged annihilation cross section of the DM particle, $m_{\mathrm{DM}}$ is the mass  of the DM particle,  $Br_i$ is the branching ratio of the annihilation channel $i$, $N_\gamma$ is the number of gamma rays produced per annihilation reaction, and $E$ is the energy. The second term is the \emph{astrophysical} (or $J$) factor, which accounts for the DM distribution and the distance of the source:
\begin{equation}
J(\dom) = \int_\dom d\Omega' \int_\mathrm{l.o.s.} \rho^2 (l, \Omega') dl~~~,
\label{eq:J}
\end{equation}
where $\rho$ is the DM density profile. The integrals run over the line--of--sight ($\mathrm{l.o.s.}$) and the observed sky region ($\dom$). Empirical estimates of DM content in dSphs, and hence the magnitudes of
expected signals  rely on inferences from stellar-kinematic data, through the Jeans equation (as closely treated  in~\cite{Charbonnier11, Martinez13}). The wide literature on DM profile evaluation suggests that this topic requires different aspects to be evaluated
when modeling galaxy DM distribution: kinematic and distribution of stars, estimated size of galaxy, in addition to the evaluation of specific stellar content, that accounts for baryons feedback. Therefore, DM profile parameterization, velocity anisotropy, and light profile modeling are needed to compute the $J$-factor and its uncertainties. Hence, the statistical uncertainties associated to the $J$-factors are due to finite sizes of stellar-kinematic data of member stars. Systematic uncertainties regard the shapes of DM density profiles as well as systematic errors can arise due to different stellar density profiles, non-spherical symmetry, and more complicated behaviors of the velocity anisotropy~\cite{Bonnivard2015}.

Annihilation of WIMPs could result in different types of gamma-ray signatures. First of all, a flux of gamma rays is expected from the $\pi^0$ decays resulting from the hadronization of SM particles produced in the DM annihilation/decay processes, and from the QCD and QED Final-State Radiation (FSR)~\cite{Cembranos13}. The resulting gamma-ray spectra are continuous, with a cutoff at the kinematical limit (i.e. at the mass of the DM particle). Other processes producing sharp, monochromatic line are in most scenarios loop-suppressed and not considered here. 

This work focuses on searching for  the DM annihilation signal. We considered the DM annihilating into the SM pairs $b\bar{b}$, $W^+W^-$, $\tau^+\tau^-$, and $\mu^+\mu^-$, employing the average gamma-ray spectrum per annihilation process ($dN_\gamma/dE$) computed for a set of DM particles of masses between 10~GeV and 100~TeV on the base of the PPPC 4 DM ID code realized on the PYTHIA simulation package version 8.135~\cite{Cirelli10}.

Since the discovery of UMaII~\cite{Zucker06}, several studies have been published on the $J$-factor estimate for this target~\cite{Martinez13,Geringer15,Bonnivard15,Hayashi16, Sanders16,Chiappo16}, all essentially confirming UMaII at the top of the ranking of highly promising dSph candidates for indirect DM searches. For our study, we use the UMaII $J$-factor parameterization as a function of the angular distance to the DM halo center given in~\cite{Geringer15}, which is largely compatible with the other determinations found in literature. In particular, in our analysis we considered the value of log$_{10}$($J$($\theta_\mathrm{max}$)~[GeV$^2$~cm$^{-5}$])=$19.42^{+0.44}_{-0.42}$ for the astrophysical factor integrated up to the maximum radius of the UMaII DM halo.

\section{MAGIC and the UMaII observation campaign}
\label{S:3}

The MAGIC (Major Atmospheric Gamma Imaging Cherenkov) telescopes are a system of two $17$~m diameter telescopes located at the Roque de los Muchachos Observatory ($28.8^\circ$ N, $17.9^\circ$ W; 2200~m a.s.l.), in the Canary island of La Palma (Spain). The two telescopes are both equipped with a fast imaging camera of $3.5^\circ$ field of view and are able to detect cosmic gamma rays in the VHE domain through the Cherenkov light produced by the atmospheric showers initiated by cosmic particles entering the Earth atmosphere. The whole MAGIC system underwent several hardware upgrades~\cite{aleksicI16} and since its latest upgrade (accomplished in summer 2014 \cite{aleksicII16}) the system has improved considerably its performance close to the energy threshold, currently providing the world-best sensitivity around $\sim$100~GeV. This represents a crucial achievement for indirect DM searches at VHE, given the typical continuum spectra expected from DM annihilation/decay processes (which makes a good sensitivity at low energy threshold a key performance factor).

UMaII was observed by MAGIC between December 2014 and April 2016, for a total of 106.8 hours. Since the observations started right after the latest upgrade of the system, the whole data sample was taken with the same hardware conditions and optimal performance. The survey was carried out in the false source tracking (or ``wobble'') mode~\cite{Formin94}, in which two wobble positions offset by $0.4^\circ$ from the center of target in opposite RA direction were alternated every 20 minutes. For each wobble pointing direction, the residual background associated to the ON region around UMaII was estimated from the (OFF) region placed at the same relative location with respect to the pointing direction of the complementary wobble observation. With this configuration, the distance between the center of the ON and OFF nominal positions in the MAGIC cameras was always kept at 0.8$^{\circ}$. The data were taken at medium zenith angles, ranging between $\sim$35$^\circ$ and $\sim$45$^\circ$, being the culmination of the source at MAGIC site at 35$^\circ$. This resulted in an analysis energy threshold (defined as the peak of the  energy distribution for a Monte Carlo simulated  Crab-Nebula like gamma-ray data set after all analysis cuts) of $\sim$120~GeV. 

\section{Standard data reduction and results}
\label{S:4}

The standard data reduction was performed with MARS~\cite{Zanin13}, the official MAGIC data reconstruction and analysis package. Data quality selection was based on LIDAR information~\cite{Fruck15}. The selection resulted in 94.8~hours of excellent-quality data. After the standard data calibration, image cleaning and parameterization, events with a total amount of signal in the recorded showers below 50~photo-electrons (for any telescope) were rejected. Then, the main stereo parameters were calculated combining the information coming from the individual telescopes. The gamma/hadron separation was achieved by means of a multivariate method called Random Forest (RF)~\cite{RF}. The algorithm employs basic image, timing and stereo parameters to compute a gamma/hadron discriminator called \emph{Hadronness} by comparison of real (hadronic-dominated) data with dedicated Monte Carlo gamma-ray simulations. The estimate of the arrival direction of the events was performed as well with the RF method, making use solely of MC gamma-ray simulations. This quantity was eventually used to compute the so-called $\theta^2$ parameter, which is the squared angular distance between the reconstructed event direction and the nominal position of the target. Finally, the energy reconstruction of the events was achieved by averaging individual energy estimators for both telescopes based on look-up tables~\cite{aleksicPerf12}. The whole standard analysis procedures was validated by means of contemporaneous Crab Nebula observations (at the same zenith range of UMaII observations), which provided the expected performance in terms of sensitivity and spectral behavior.

The analysis cuts were optimized by means of a dedicated procedure aimed at finding the best sensitivity\footnote{Here, ``sensitivity'' is defined as the average upper limit that would be obtained by an ensemble of experiments with the expected background and no signal~\cite{Feldman97}.} to the thermally averaged annihilation cross section $\langle\sigma_\mathrm{ann}v\rangle$ of the full likelihood analysis (see Sec.~\ref{S:5}) as a function of different cuts in Hadronness and $\theta^2$ parameters. The extension of the source was taken into account in the optimization of the cuts as explained in Sec~\ref{S:5}. 

Defining the profile likelihood ratio $\lp$~\cite{Olive14} as a function of $\langle \sigma \mathit{v} \rangle$ for the measured dataset $\mathcal{D}$:
\begin{equation}
 \lp(\sv |\mathcal{D})=\frac{\lkl(\sv;\hat{\hat{\nu}}|\mathcal{D})}{\lkl({\widehat{\sv};\hat{\nu}|\mathcal{D}})}~~~,
\label{eq:lklratioprofile}
\end{equation}
where $\lkl$ is the likelihood function (whose detailed expression is  the Eq.~\ref{eq:binnedlikelihood}), depending on the nuisance parameters $\nu$ (i.e.  the ratio of exposures between the OFF and ON regions and the expected number of background events in the OFF region, see next section), in particular $\hat{\nu}$ and $\widehat{\sv}$ are the values maximizing it, and the $\hat{\hat{\nu}}$ maximizes  $\lkl$ for a given value of $\sv$.  The  sensitivity can be approximated by  $\svsvt=\langle \sigma \mathit{v} \rangle_{2.71} - \widehat{\langle \sigma \mathit{v} \rangle}$, where  $\langle \sigma \mathit{v} \rangle_{2.71}$ is defined by $-2\ln\lp(\langle \sigma \mathit{v} \rangle_{2.71}|\mathcal{D})=2.71$ (for a detailed explanation, see~\cite{ahnen16}).
As result of the optimization cuts procedure, the optimal cuts  $\theta=0.3^\circ$\footnote{Due to the extended estimated size of the UMaII DM halo ($\theta_\mathrm{max} \simeq 0.53^\circ$), we evaluated that the leakage of the signal into the chosen OFF region of $0.3^\circ$ is less than a thousandth of the total signal and, thus, negligeble.} and Hadronness retaining 70\% MC gamma rays~--~independently in 40 logarithmic energy bin cuts between 10~GeV and 100~TeV~--~provided the best choice for the gamma/hadron separation cuts for all considered final states and two benchmark DM masses: 1~TeV and 10~TeV (i.e. where the most stringent constraints can be typically achieved).

\begin{figure} [htb!]
\begin{center}
\includegraphics[height=8cm]{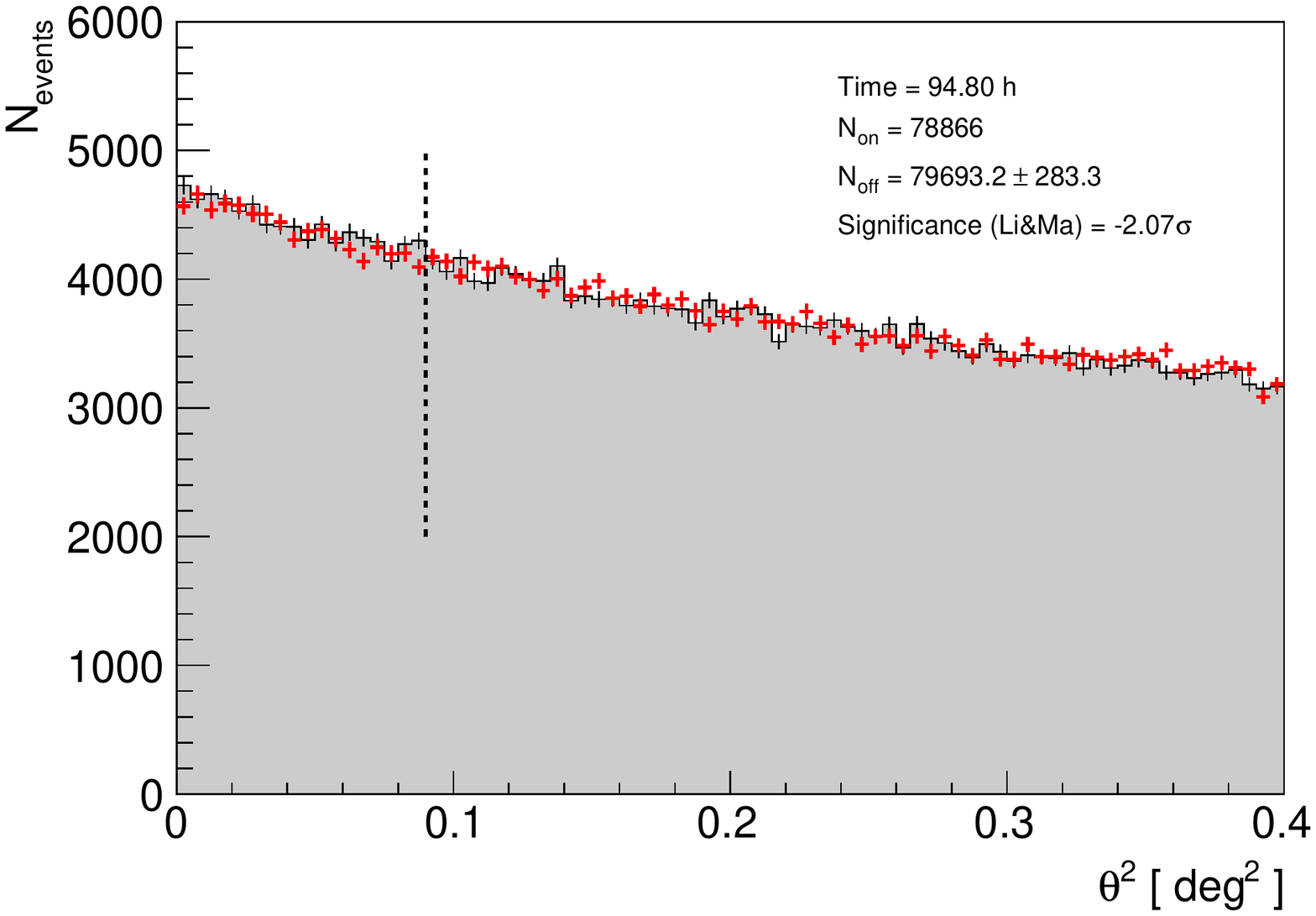}
\end{center}
\caption[]
{
$\theta^{2}$ distributions of  ON (red) and OFF (grey) regions resulting from 94.8~hours of MAGIC stereoscopic observations of UMaII taken between December 2014 and April 2016, with an energy threshold of 120~GeV. The region between zero and the vertical dashed line (at $\theta^2=$0.09$^{\circ 2}$) represents the ON and OFF integration regions.
} 
\label{fig2}
\end{figure}

The overall search for a gamma-ray signal from UMaII was performed with the so-called $\theta^{2}$-plot, after the application of the energy-dependent optimized cuts, and within the chosen integration $\theta^{2}$ region. In order to evaluate the residual background of the observation, the $\theta^{2}$ distribution around a nominal background control region was also calculated. \autoref{fig2} shows the resulting $\theta^{2}$-plot. No significant gamma-ray excess~\footnote{The deficit found in the $\theta^{2}$-plot ($\simeq$-2$\sigma$) is at the level of 1\% of the background, i.e. within the systematic effect of 1.5\% properly taken into account in the likelihood analysis (see \autoref{S:5}).} was found around the nominal position of UMaII.

\begin{figure} [htb!]
\begin{center}
\includegraphics[height=8cm]{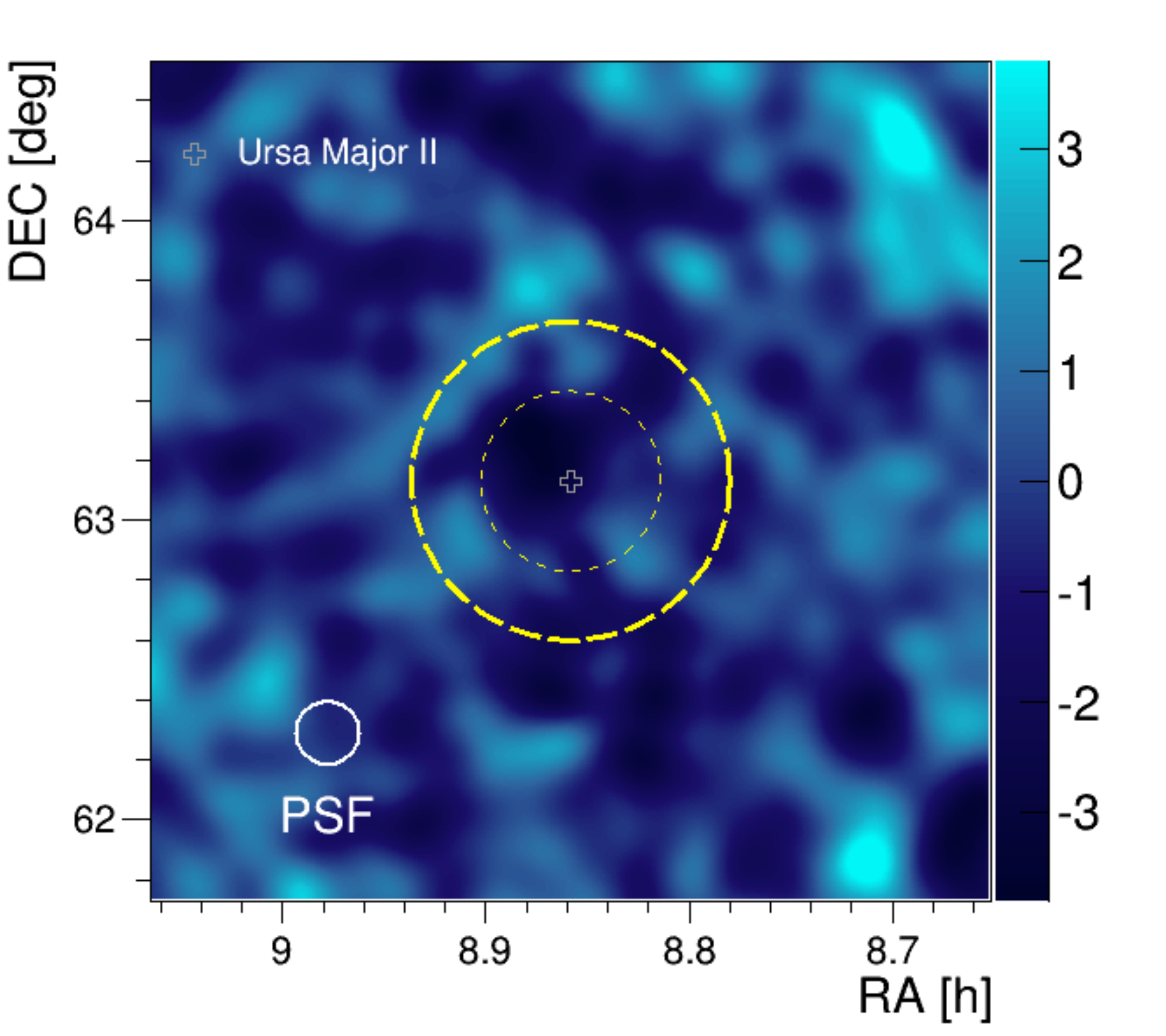}
\end{center}
\caption[]
{
Significance sky-map centered at the UMaII sky position from 94.8~hours of MAGIC stereoscopic observations taken between December 2014 and April 2016, with an energy threshold of 120~GeV. The UMaII center position is marked with an empty white cross. The color scale represents the test statistic value distribution. The dashed yellow (external) circle represents the region within the maximum-radius of $0.53^{\circ}$ of the UMaII DM halo. The dotted yellow (internal) circle represents the region within the optimized analysis $\theta$ cut of $0.3^{\circ}$. The MAGIC PSF (for the given analysis cuts) of $0.11^{\circ}$ is also shown (white circle).
}
\label{fig3}
\end{figure}

In \autoref{fig3} the sky-map~\cite{Aleksic11,Lombardi11} centered in the target sky position calculated with the application of the same analysis cuts is depicted. Our test statistic  is taken from~\cite{LiMa83} (Eq. 17), applied on a smoothed and modelled background estimation. The statistical hypothesis tested, that is the null hypothesis, mostly resembles a  $\chi^2$ distribution. This circumstance allows to apply the Wilks theorem~\cite{Wilks38} in order to estimate the level of agreement between the data and the  hypothesis, inferring the significance of the observed result.
Also in this case, no significant gamma-ray excess over the background in the sky region of UMaII DM halo (yellow dashed circle) was found.

Since no hint of a gamma-ray excess was found, the analysis of UMaII data proceeded with the computation of the constraints on the DM annihilation cross-section for different considered channels, using the full likelihood analysis method. In the next sections, before presenting the final results achieved in this work, we discuss the analysis method and its input quantities.

\section{Full likelihood analysis method}
\label{S:5}

The full likelihood allows the exploitation of spectral features of the expected DM signal to optimize the sensitivity with respect to a conventional analysis approach. A detailed review of this method and its formalism can be found in~\cite{aleksic12}. 

The likelihood $\lkl$ is a function depending on the expected number of gamma-rays $g$ detected as a function of the estimated energy $E'$ and an observation time $T_\mathrm{obs}$. In addition, the likelihood depends on several nuisance parameters (for a detailed explanation of the likelihood function see~\cite{ahnen16}). In this study we performed a binned analysis, i.e. we considered $N_\mathrm{bins}$ bins in estimated energy in the full likelihood function introduced in \cite{aleksic12} (and used in the previous MAGIC DM studies \cite{aleksic14,ahnen16}). Here $\lkl$ is the product of two likelihood functions $\lkl_i$, one for each set of data taken in the two different wobble pointing directions ($i$).  The binned version reads as:
\begin{eqnarray}\label{eq:binnedlikelihood}
\lkl_i (\sv;\bm{\nu}_i\, |\, \data_i)  & = &
\lkl_i (\sv; \{\bij\}_{j=1,\dots,\Nbins}, J , \tau_i\, |\, 
 ( \Nonij,\Noffij)_{j=1,\dots,\Nbins} \nonumber )\\
&=&
\prod_{j=1}^{\Nbins} 
\biggl[ \frac{(\gij(\sv) +\bij)^\Nonij}{\Nonij!} e^{-(\gij(\sv)+\bij)} \nonumber \\
& & \times\,
\frac{(\tau_i \bij)^\Noffij}{\Noffij!} e^{-(\tau_i\bij)}\biggr]\\
& & \times\, 
\mathcal{T}(\tau_i|\tauobsi,\sigmataui) \times \mathcal{J}(J|\Jobs,\sigma_{\log_{10}J}) \nonumber~~~,
\end{eqnarray}
where the index $i=1,2$. The $\bm{\nu}_i$ represents the nuisance parameters and $\data_i$ the dataset; $\gij$, $\bij$ and $\Nonij$ are the estimated number of signal and background events, and the number of observed events, respectively, in the $j$-th ON energy bin; $\Noffij$ is the number of observed events in the corresponding OFF bin; $\mathcal{J}$ is the likelihood for the $J$-factor, $\mathcal{T}$ is the likelihood for $\tau_i$ (the OFF/ON acceptance ratio), determined from the ratio of the number of observed events in regions adjacent to the OFF and ON regions, parameterized by a Gaussian function with mean $\tauobsi$ and variance $\sigmataui^2$, which include statistical and systematics uncertainties. In the present analysis, we considered a  systematic uncertainty of $\sigma_{\tau_\mathrm{syst}}=1.5\%$ on the estimate of the residual background (see the \autoref{tab:tauValues}). This  value has been established on the base of a dedicated performance study~\cite{aleksicII16}. At high statistics ($>10^4$ ON events, corresponding to $\sim$50h), the systematic uncertainty dominates and is due to the possible difference in camera acceptance between the ON and OFF regions. $\bij$, $J$ and $\tau_i$ are  nuisance parameters, whereas $\gij$ depend on the free parameter $\sv$ through:
\begin{equation}\label{eq:freeparameter}
\gij(\sv) = \Tobsi \int_\Eminj^\Emaxj dE'
\int_0^\infty dE
\frac{d\Phi(\sv)}{dE}\, \aeff(E)\, G(E'|E)~~~,
\end{equation}
where $\Tobsi$ is the total observation time, $E$ and $E'$ the true and estimated gamma-ray energy, respectively, and $\Eminj$ and $\Emaxj$ the minimum and maximum energies, respectively, of the $j$-th energy bin. Finally, $\aeff$ is the effective collection area and $G$ the probability density function (PDF) of the energy estimator, both computed from a Monte Carlo simulated gamma-ray dataset following the spatial distribution expected for DM-induced signals from UMaII (see appendix~\ref{A:1} for further details).
\begin{table}
\centering
\begin {tabular}{|c|c|c|c|c|}
\hline
Wobble position&  Eff ON time & $\tau$ & $\sigma_{\tau_\mathrm{stat}}$ & $\sigma_{\tau_\mathrm{syst}}$\\
               & [h] &  &  &\\
\hline
W1        & $49.29$  & $0.9111$ & $0.0037$ & $0.0137$ \\
\hline
W2  & $45.49$  & $1.0943$ & $0.0045$ & $0.0164$\\
\hline
\end{tabular}
\caption{Effective observation time (\emph{second column}), ON/OFF acceptance ratio $\tau$ (\emph{third column}), statistic error for $\tau$ (\emph{fourth column}), systematic error for $\tau$ (\emph{last column}) considered in this analysis, for both wobble pointing positions (\emph{first column}).}
\label{tab:tauValues}
\end{table}

The input of the likelihood are the number of events detected in the ON and OFF regions for the different bins in estimated energy --after proper cuts in Hadronness and $\theta^2$ parameter-- as well as the instrument response functions (IRFs) computed for the specific observation period and the extension of the source.

UMaII is an extended source for the MAGIC PSF ($\sim0.1^\circ$), being $\theta_\mathrm{max}=0.53^\circ$ and the ``half-light-radius'' equal to $\theta_{0.5}=0.24^\circ$. For this reason, in order to take into acount the extension of DM emission region, the IRFs were computed from MC simulations following UMaII morphology (see details in appendix~\ref{A:1}).

Using the profile likelihood ratio $\lp$ (see Eq.~\ref{eq:lklratioprofile}) we test hypotheses that assume the flux computed with Eq.~\ref{eq:particlephysics} and~\ref{eq:J}, considering ``pure'' annihilation channels: $b\bar{b}$, $\tau^+\tau^-$,  $\mu^+\mu^-$, and $W^+W^-$; one-sided 95\% confidence level (CL) limits are given by the largest of the two $\svdsu$ solutions (as defined in section~\ref{S:4}).

\section{Results on dark matter annihilation models}
\label{S:6}

In this section we present the 95\% CL upper limits on the thermally-averaged cross-section $\langle\sigma_\mathrm{ann}v\rangle$ for DM particles  annihilating with 100\% branching ratio into different SM particle pairs achieved in 94.8~hours of selected data of the UMaII campaign. The search was performed for DM particles of masses between 100~GeV and 100~TeV for annihilation scenarios. In our full likelihood approach, we followed the same prescription adopted in~\cite{ahnen16}, restricting the value of $\langle\sigma_{ann}v\rangle$ to the physical ($\geq$~0) region. Furthermore, no additional boosts, either from the presence of substructures~\cite{Strigari07} or from quantum effects~\cite{Hisano05}, were assumed for computing the final results.

\begin{figure} [htb!]
\begin{center}
\includegraphics[height=7cm]{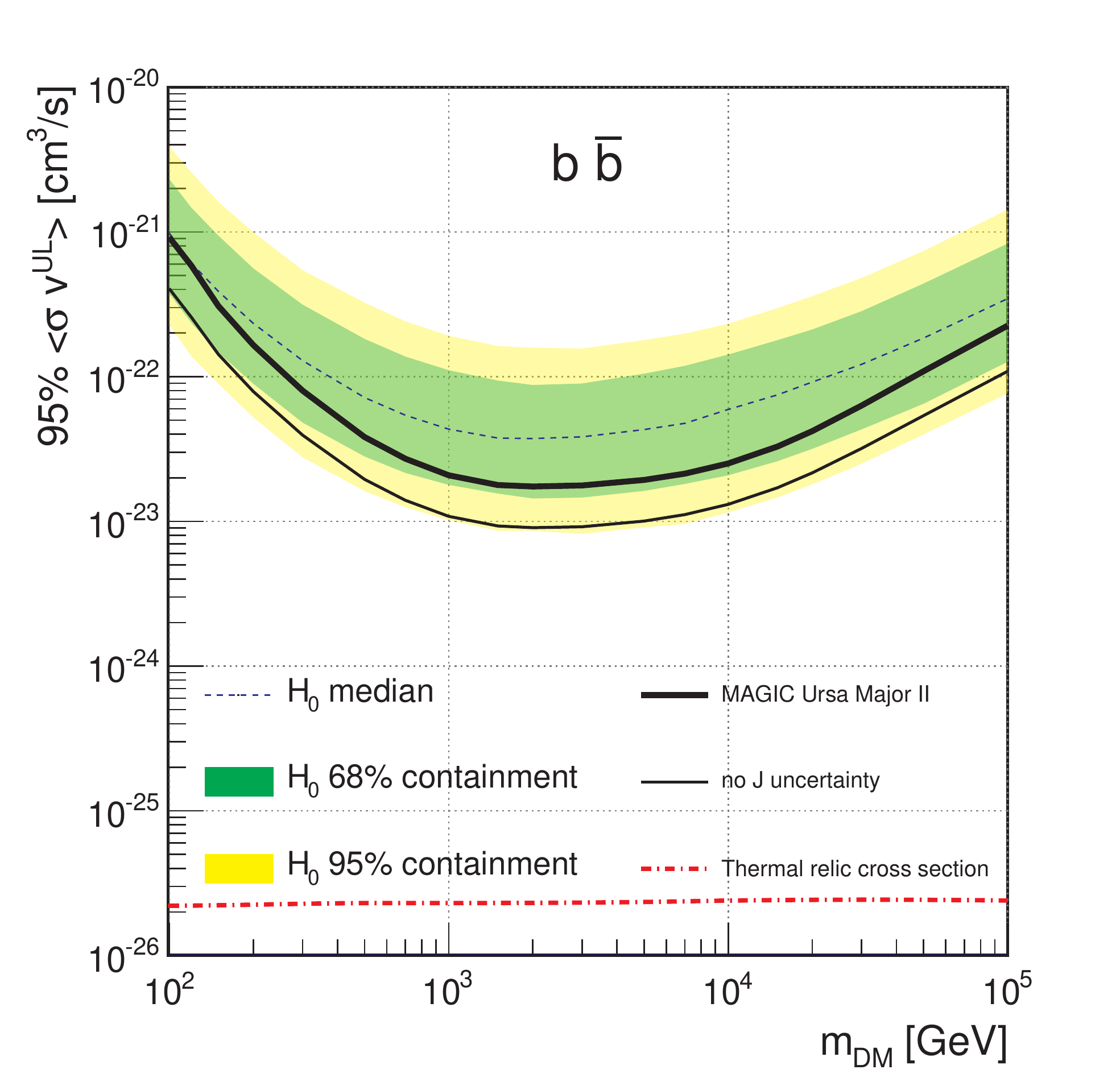}
\includegraphics[height=7cm]{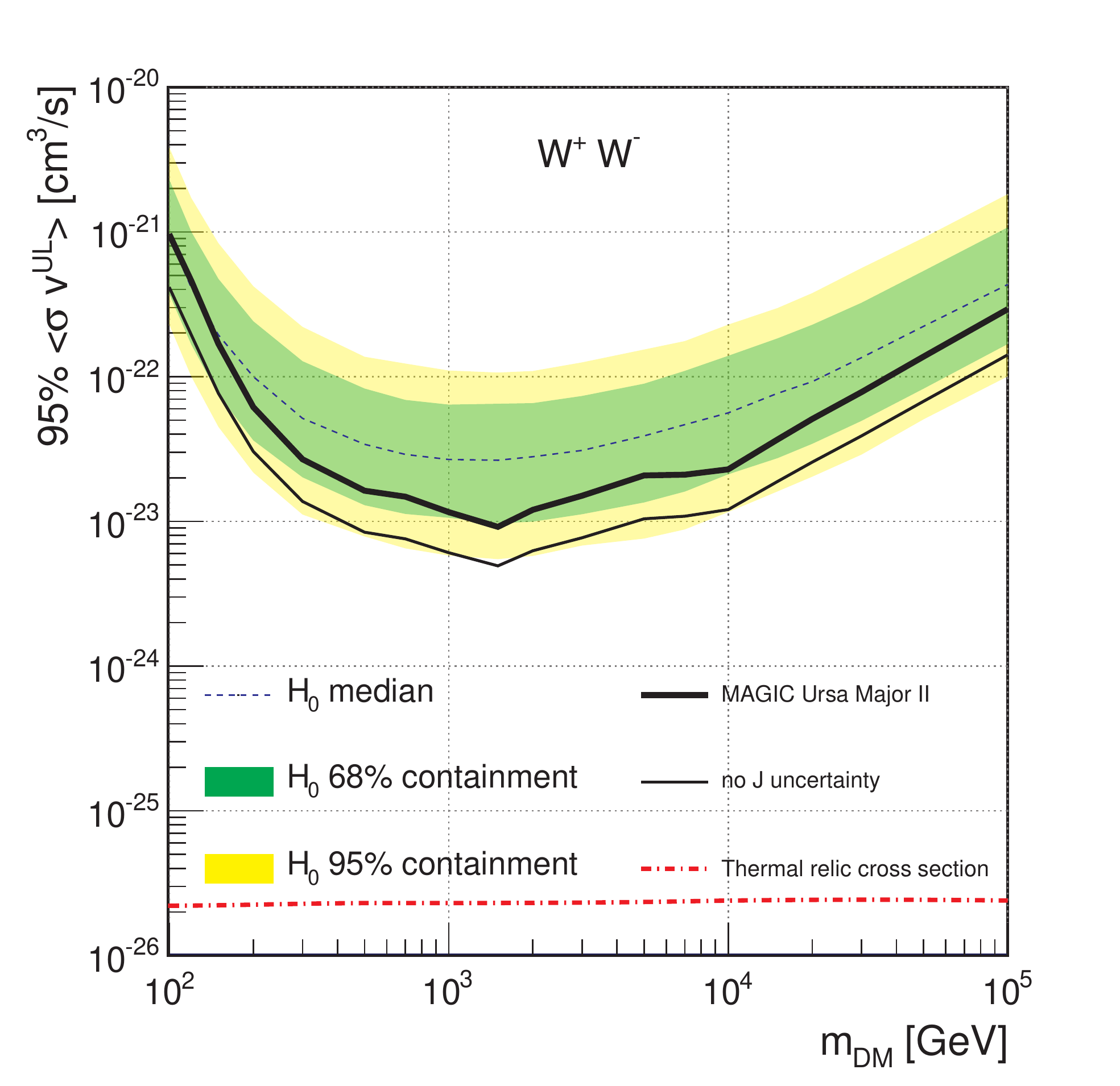}
\includegraphics[height=7cm]{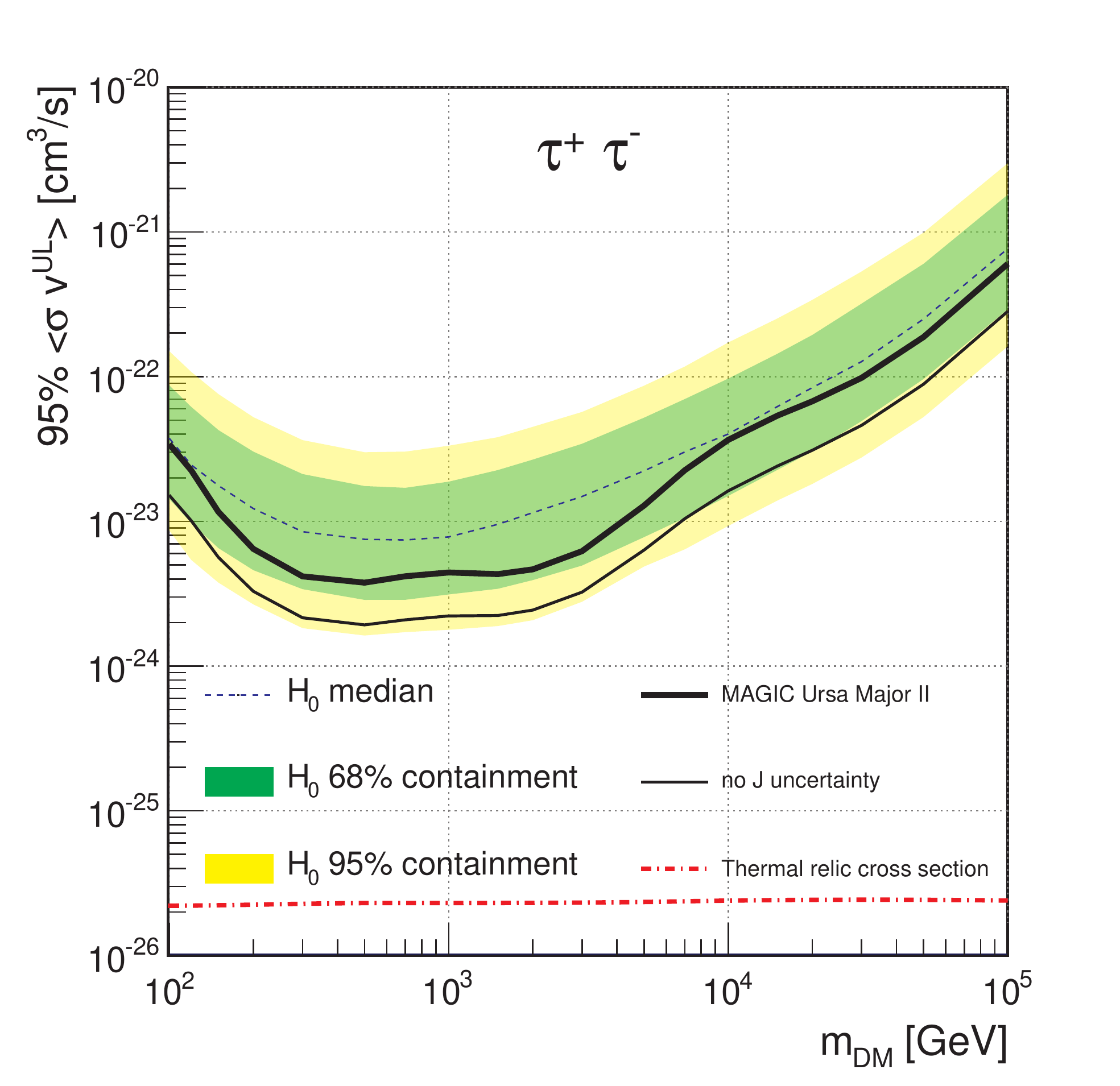}
\includegraphics[height=7cm]{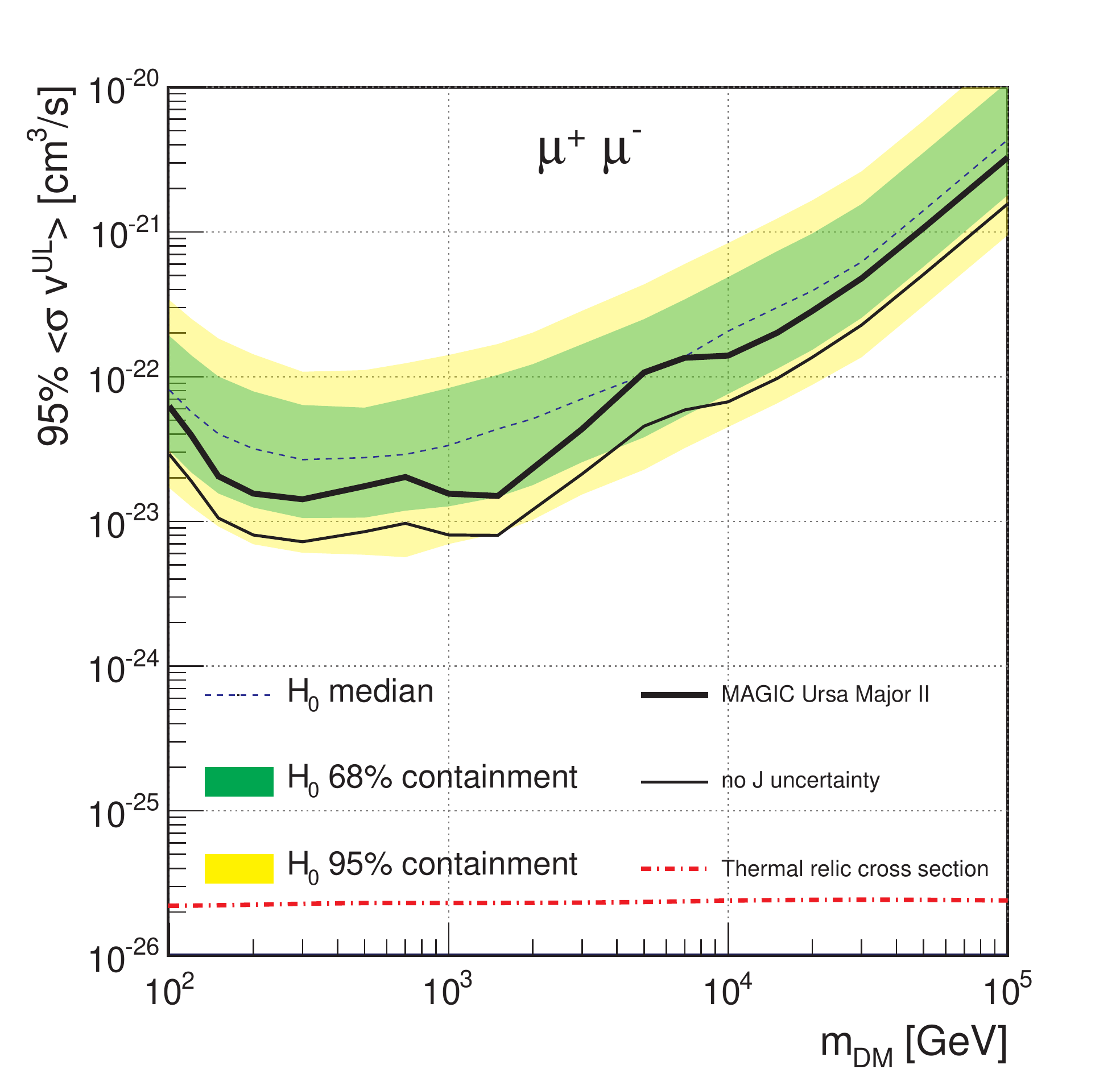}
\end{center}
\caption[]
{
95\% CL upper limits on $\langle\sigma_\mathrm{ann}\mathit{v}\rangle$ for DM particles annihilating into $b\bar{b}$ (\emph{upper-left}), $W^+W^-$ (\emph{upper-right}), $\tau^+\tau^-$ (\emph{bottom-left}) and $\mu^+\mu^-$ (\emph{bottom-right}) pairs. Thick-solid and thin-solid lines show, respectively, the limits obtained with 94.8~h of UMaII observation, considering the $J$-factor a nuisance parameter and fixing its value in the likelihood. The thin-dotted line, green and yellow bands show, respectively, the median and the symmetrical, two-sided 68\% and 95\% containment bands for the distribution of limits under the null hypothesis. The red-dashed-dotted line shows the thermal relic cross-section from~\cite{Steigman12}.
}
\label{fig4}
\end{figure}

In \autoref{fig4}, the 95\% CL upper limits on $\langle\sigma_\mathrm{ann}\mathit{v}\rangle$, for DM particles annihilating into $b\bar{b}$, $W^+W^-$, $\tau^+\tau^-$, and $\mu^+\mu^-$, achieved after the application of the optimized cuts and with a binned ($N_\mathrm{bins}=30$) likelihood analysis are shown. In addition, the two-sided 68\% and 95\% containment bands for the distribution of limits under the null hypothesis are also reported. The containment bands were computed from the distribution of the upper limits obtained from the analysis of 1000 realizations of the null hypothesis ($\sv=0$), consisting of fast simulations (for both ON and background regions) generated from background PDFs, assuming similar exposures as for the real data, and $J$-factors assumed as nuisance parameters in the full likelihood function.
All bounds are consistent with the no-detection scenario. The achieved results represent among the most stringent and robust constraints to the annihilation cross-section obtained from observations of single dSphs, in the TeV mass region.  In particular, our strongest limit (95\% CL) corresponds to a $\sim$0.5~TeV DM particle annihilating into $\tau^+\tau^-$, and is of order $\langle\sigma_\mathrm{ann}\mathit{v}\rangle \simeq 3.8 \times 10^{-24}$ cm$^{3}$ s$^{-1}$. The results are comparable with the recently published limits achieved by VERITAS Collaboration in the  joint analysis of data collected on four dSphs (for a total of  216~hours of collected data)~\cite{Veritas17} and the combined results of HESS campaigns on dSphs (including 5 dSphs for an amount of 140~hours)~\cite{Abramowski14}.

Due to the difference in the analysis method developed and adopted for the UMaII data with respect to the previous dSph campaigns, mainly in the  treatment of the nuisance parameter \emph{J} and in the background modeling, a straightforward comparison with previous MAGIC results is not easily achievable. Nevertheless, the results shown in the present work are comparable (within a factor of $\sim3$) with those obtained with the Segue~1 campaign ($\sim~160$~hours)~\cite{aleksic14}, once the difference in the targets' exposure, the treatment of the systematics for $\tau$, and other differences in the analyses are taken into account. We reserve for future publications to combine all MAGIC data collected on dSphs in a new homogeneous analysis, that will take advantage of the optimized tools tested and used in the present work.

\section{Summary and Conclusions}
\label{S:7}

The MAGIC telescopes conducted a deep observation campaign at VHE toward the UMaII dSph, a promising target for indirect DM searches. The source was observed between December 2014 and April 2016, resulting in 94.8~hours of excellent-quality data. This campaign represents an important step toward an optimal ``target diversification strategy'' at VHE aimed at enhancing the chances of discovery of DM signals and reducing possible biases in target selection.

Since no significant gamma-ray excess was found in the UMaII data, the observations were used to derive the constraints  to the annihilation cross-section assuming annihilation into the SM pairs $b\bar{b}$, $W^+W^-$, $\tau^+\tau^-$, and $\mu^+\mu^-$, for DM particles in the 100~GeV--100~TeV mass range. The 95\% CL limits obtained in this work, by means of the full likelihood analysis method, are among the most stringent and robust achieved so far from observations of dSphs at the TeV mass scale. For the first time we optimized the DM search in dSph exploiting the morphology information of the target by taking into account the extension of the UMaII DM halo (see appendix~\ref{A:1}).

Since the beginning of the UMaII campaign with MAGIC, in the last two years new interesting dSphs for DM searches have been discovered. In this respect the MAGIC program of DM search in dSphs continues, following the target diversification strategy proposed with the UMaII campaign. Moreover, thanks to the full likelihood analysis method, the results of this work will have a natural development in a more general framework of joint analysis involving different dSphs and (possibly) different instruments.

\acknowledgments

We would like to thank the Instituto de Astrof\'{\i}sica de Canarias for the excellent working conditions at the Observatorio del Roque de los Muchachos in La Palma. The financial support of the German BMBF and MPG, the Italian INFN and INAF, the Swiss National Fund SNF, the ERDF under the Spanish MINECO (FPA2015-69818-P, FPA2012-36668, FPA2015-68378-P, FPA2015-69210-C6-2-R, FPA2015-69210-C6-4-R, FPA2015-69210-C6-6-R, AYA2015-71042-P, AYA2016-76012-C3-1-P, ESP2015-71662-C2-2-P, CSD2009-00064), and the Japanese JSPS and MEXT is gratefully acknowledged. This work was also supported by the Spanish Centro de Excelencia ``Severo Ochoa'' SEV-2012-0234 and SEV-2015-0548, and Unidad de Excelencia ``Mar\'{\i}a de Maeztu'' MDM-2014-0369, by the Croatian Science Foundation (HrZZ) Project IP-2016-06-9782 and the University of Rijeka Project 13.12.1.3.02, by the DFG Collaborative Research Centers SFB823/C4 and SFB876/C3, the Polish National Research Centre grant UMO-2016/22/M/ST9/00382 and by the Brazilian MCTIC, CNPq and FAPERJ. The work of the author M. Vazquez Acosta is financed with grant RYC-2013-14660 of MINECO.



\appendix
\section{DONUT Monte Carlo method}
\label{A:1}

Instrument response functions (IRFs) of Cherenkov telescopes are
usually evaluated by means of Monte Carlo (MC) simulations. For many
practical purposes, it is enough to evaluate IRFs for point-like
gamma-ray sources. However, IRFs depend in general on the relative
arrival direction of the gamma ray with respect to the telescope
pointing direction. This means that the evaluation of IRFs for
extended sources of arbitrary shape would in principle need a
simulation of a gamma-ray sample with arrival directions distributed
following the particular source morphology. Such morphology is
expected to be very different from source to source (e.g.: the diffuse
emission of the Milky Way plane ~\cite{Ackermann:2017hri,Ahnen:2017crz}, 
nearby supernova remnants ~\cite{Flinders:2015baa,Aliu:2014rha} or
the expected gamma emission from dark matter halos 
~\cite{aleksic14,Palacio:2015nza}). In order to
compute the IRFs applicable to the study of these sources, while making
an efficient use of the computing resources devoted to MC simulations,
we have developed a method, which we dub donut MC, described and characterized in this appendix.

\begin{figure}
  \begin{center}
  \begin{subfigure}{.5\textwidth}
    \centering
    \includegraphics[width=1.\linewidth]{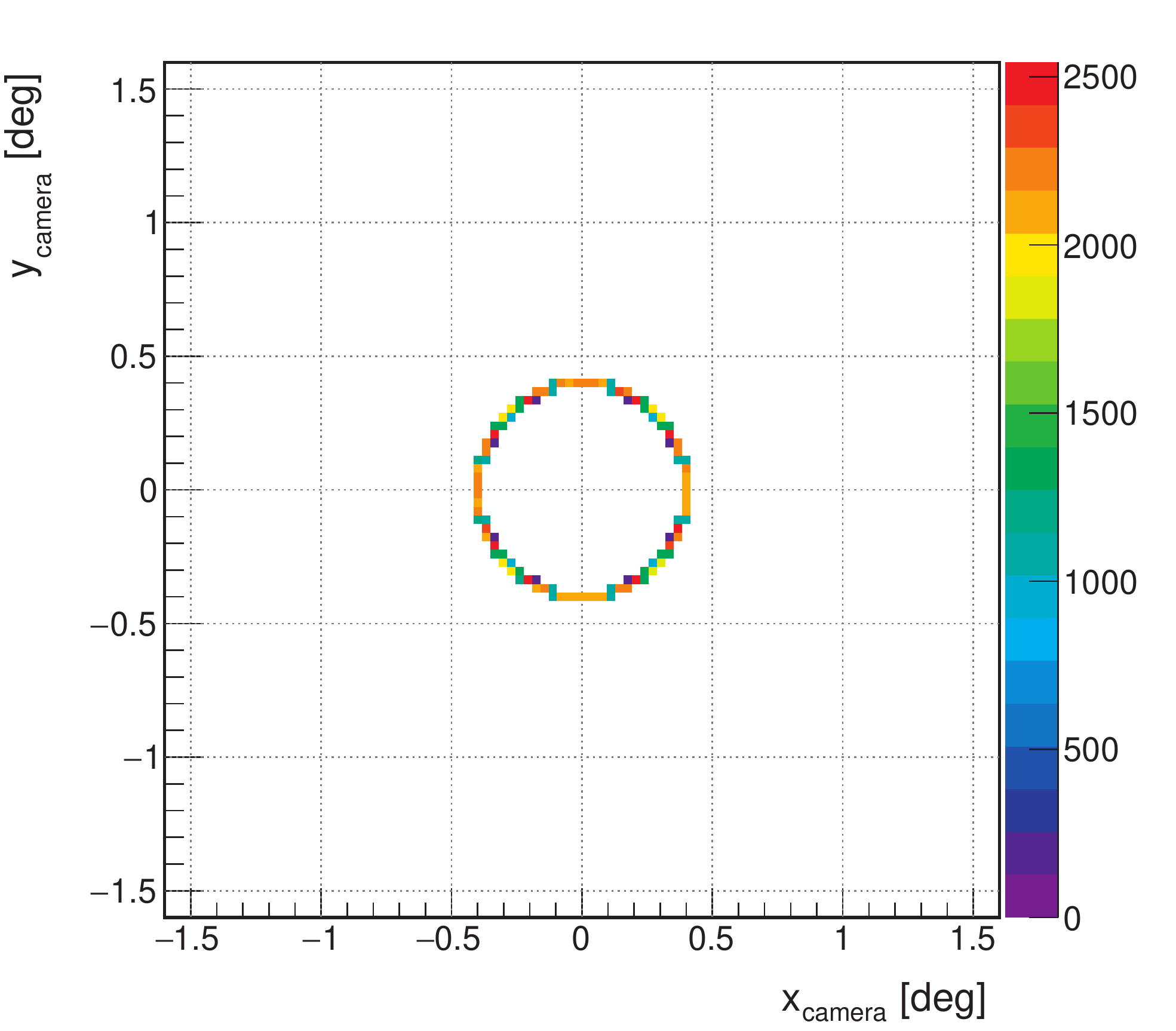}
  \end{subfigure}%
  \begin{subfigure}{.5\textwidth}
    \centering
    \includegraphics[width=1.\linewidth]{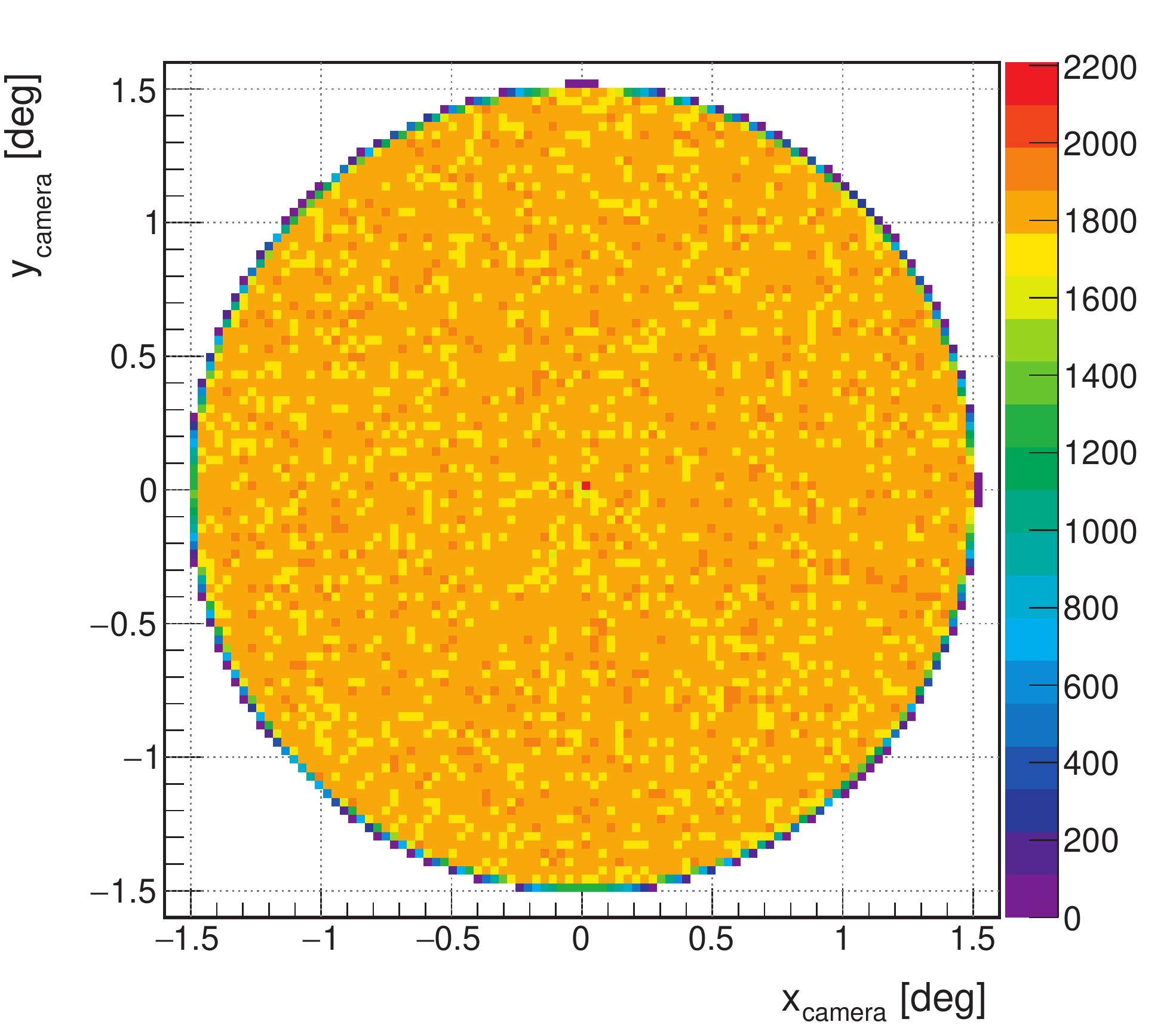}
  \end{subfigure}
  \end{center}
  \caption{Distribution of true directions of simulated events in the
    point like (left) and diffuse (right) MAGIC MC productions, shown in
    camera coordinates.}
  \label{fig:TrueDirectionMonteCarloMAGIC}
\end{figure}

\begin{figure}
  \begin{subfigure}{.25\textwidth}
    \centering
    \includegraphics[width=.95\linewidth]{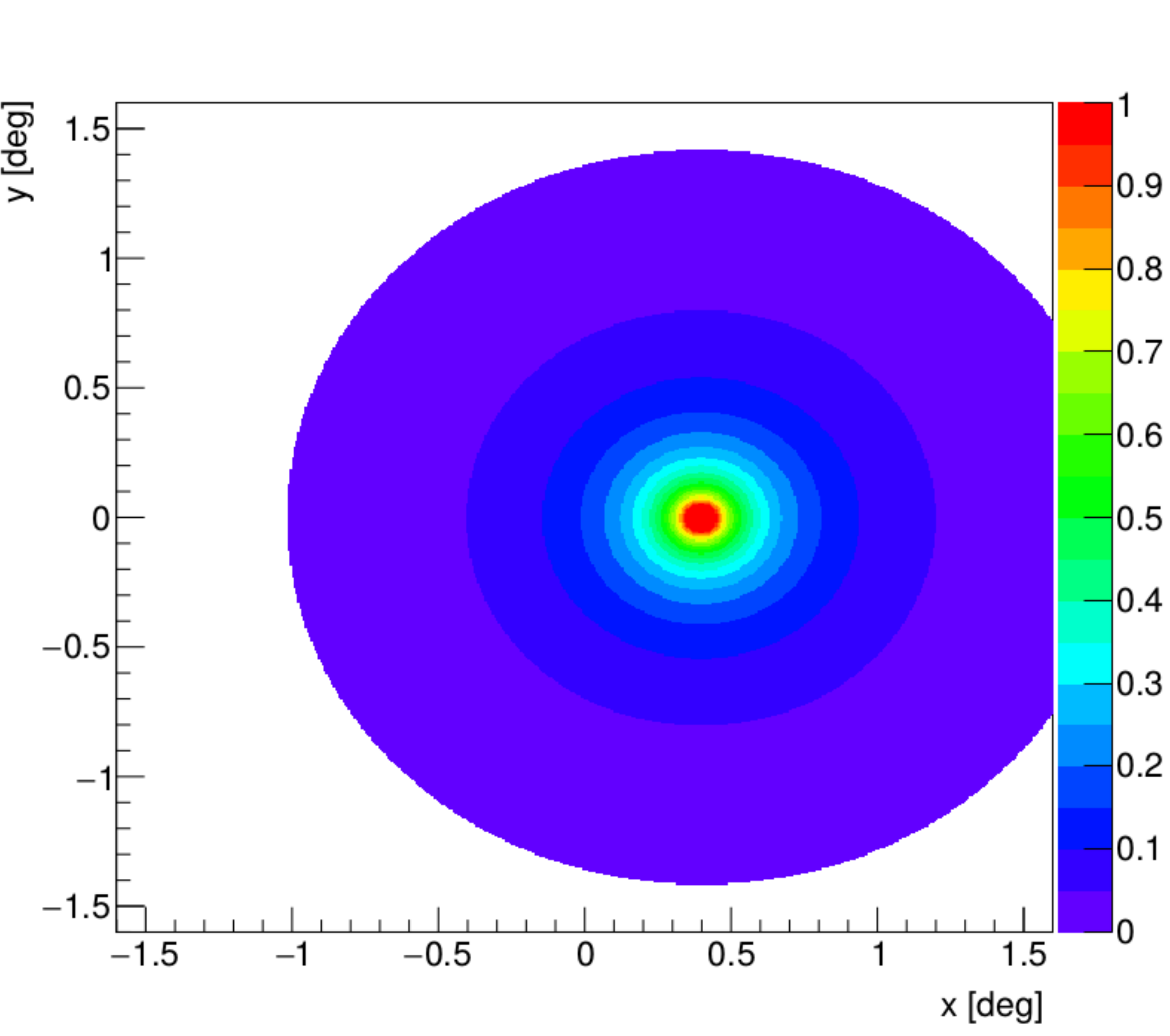}
    \label{fig:DonutMonteCarloPerseusDecay1Halos}
  \end{subfigure}%
  \begin{subfigure}{.25\textwidth}
    \centering
    \includegraphics[width=.95\linewidth]{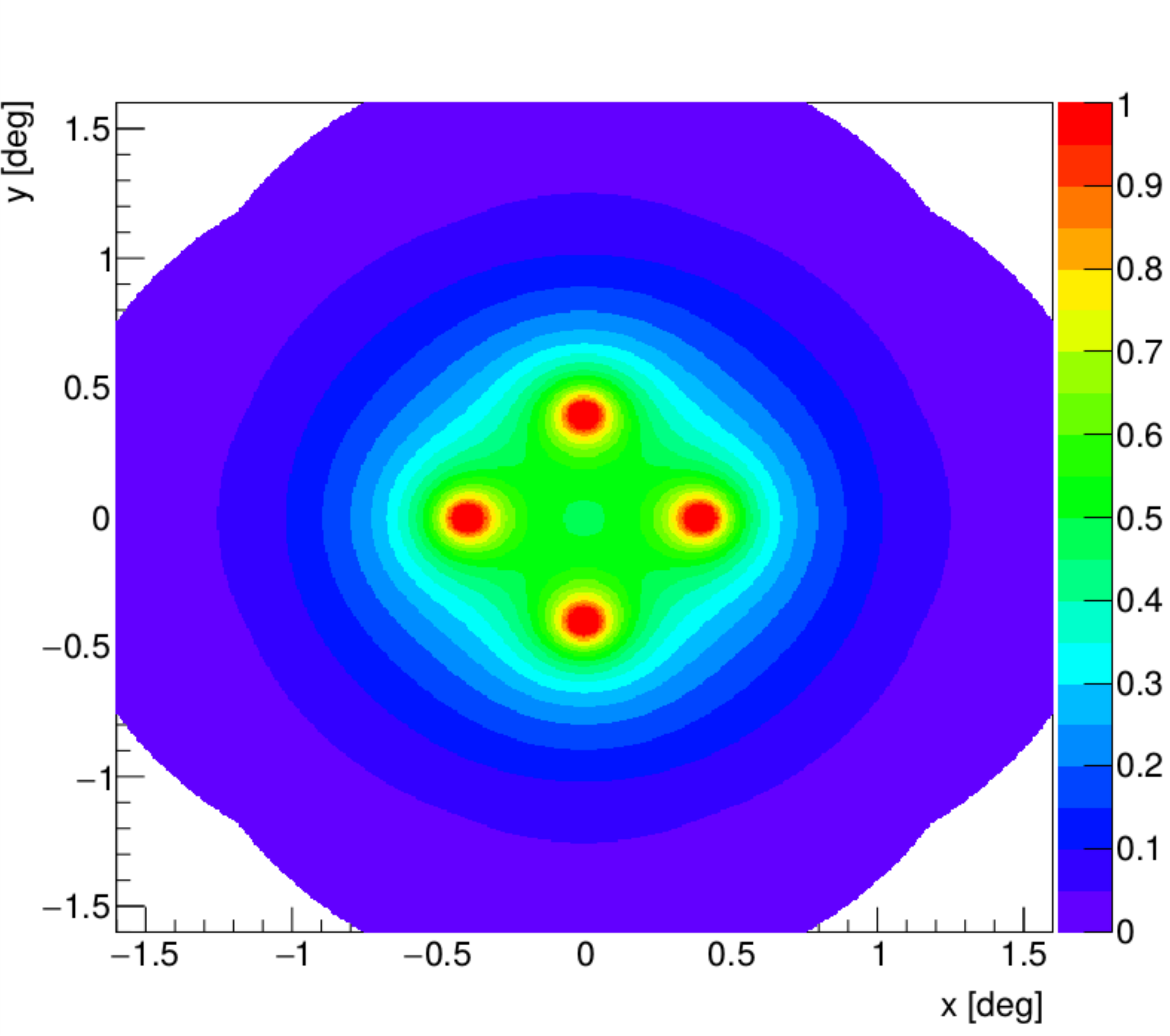}
    \label{fig:DonutMonteCarloPerseusDecay4Halos}
  \end{subfigure}%
  \begin{subfigure}{.25\textwidth}
    \centering
    \includegraphics[width=.95\linewidth]{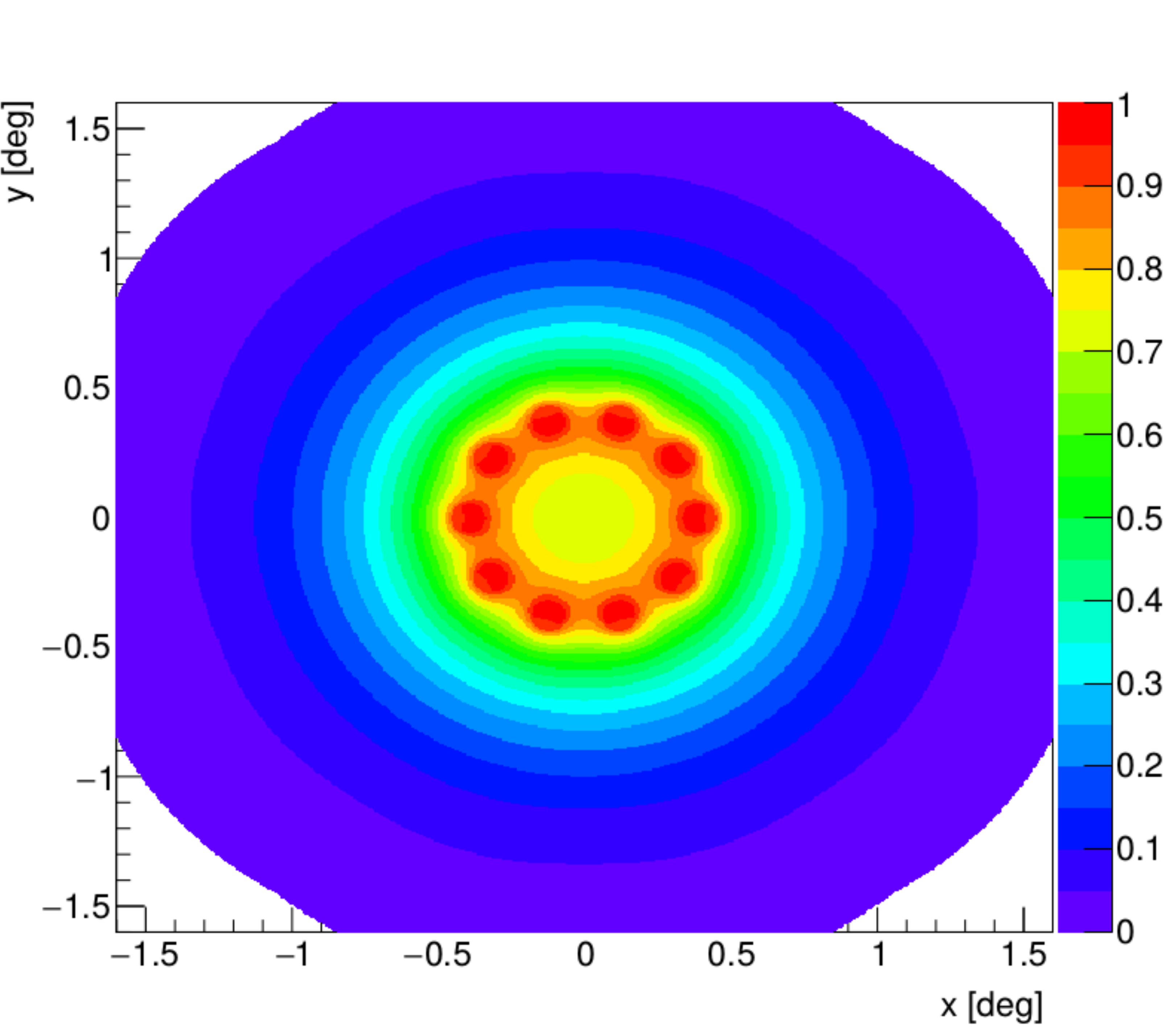}
    \label{fig:DonutMonteCarloPerseusDecay10Halos}
  \end{subfigure}%
  \begin{subfigure}{.25\textwidth}
    \centering
    \includegraphics[width=.95\linewidth]{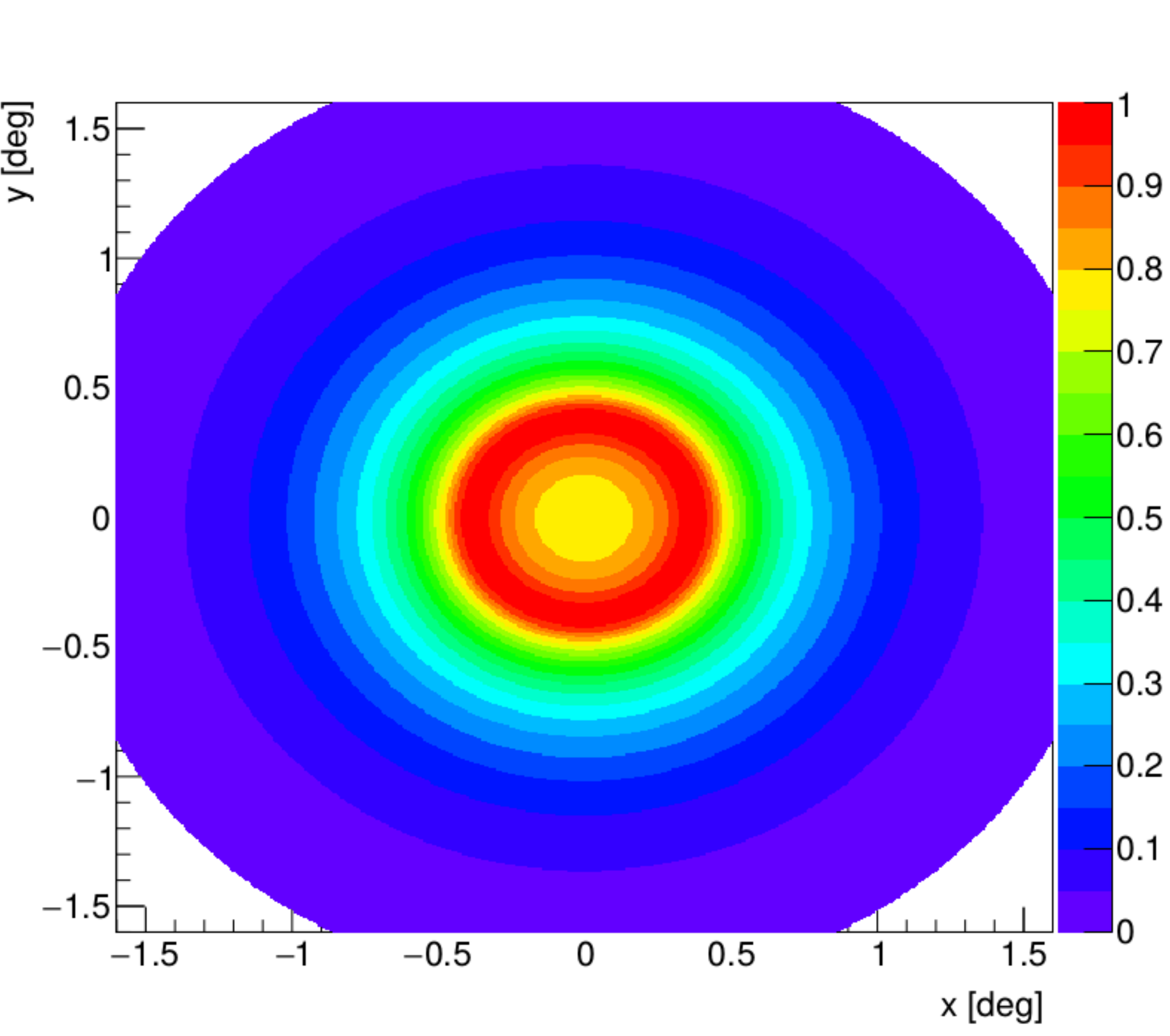}
    \label{fig:DonutMonteCarloPerseusDecay100Halos}
  \end{subfigure}
  \caption{Expected distribution of gamma-ray true directions for
    different number of source center/pointing direction orientations
    (from left to right: halo1, halo4, halo10 and halo100) realizations 
    shown in camera coordinates, for a given typical radially symmetric source.
    Image in the most right corresponds to the donut method joint PDF 
    (see text for further explanations).}
  \label{fig:DonutMonteCarloPerseusDecay}
\end{figure}

MAGIC observations of point-like sources are carried out in wobble
mode, i.e. with the telescope pointing successively at two or more
directions 0.4$^{\circ}$ away from the source position. The corresponding
IRFs are computed using the so-called \emph{point-like MC}, which
consists of gamma rays simulated with true directions uniformly
distributed in a ring centered at the telescope pointing direction and
a radius of 0.4$^{\circ}$ (see \autoref{fig:TrueDirectionMonteCarloMAGIC}, left) 
to cover all possible
orientations between the pointing direction and the source position.
Although extended sources, on the other hand, do not have a well
defined source position, the wobbling procedure is still applied by
pointing the telescope 0.4$^{\circ}$ away from a certain direction that we
call the \emph{source center}. For evaluating the IRFs in this case,
the natural procedure would be to simulate gamma rays with true
directions following the source morphology around the source center,
and the source centers uniformly distributed in a ring centered at the
telescope pointing direction and a radius of 0.4$^{\circ}$. Such dedicated
MC production would demand at least as much computer resources as the
point-like production, but would only be aplicable for the study of a
very specific source morphology. As an effective alternative, we have
developed a method to select simulated events from a MC production
consisting of gamma rays with true directions uniformly distributed in
a 1.5$^{\circ}$ radius FoV (called \emph{diffuse MC}, see 
\autoref{fig:TrueDirectionMonteCarloMAGIC}, right). This procedure only
adds a negligible overhead to the computing-intensive process of the
full diffuse MC production, which is common to all possible source
morphologies, thus making an efficient use of the computing resources
available to MC simulations. For the case of
a moderately-extended, radially symmetric source, the distribution of
true gamma-ray directions resulting from our procedure has
the shape resembling that of a 
donut 
(see \autoref{fig:DonutMonteCarloPerseusDecay}, right), where
the name of the method comes from. The rest of this appendix briefly describes the
procedure of donut MC selection and of the consistency tests that show
that our implementation actually produces the expected results.
The donut MC method is the procedure by which we produce a MC
sample, specific for the study of given source morphology, by
selecting events from the diffuse MC (see \autoref{fig:TrueDirectionMonteCarloMAGIC} 
right and \autoref{fig:DonutMonteCarloPerseusDecay} right). 
The method maximizes the number of selected events in the new
MC sample, while keeping them statistically uncorrelated. 
\\

\begin{figure}
  \begin{center}
  \begin{subfigure}{.5\textwidth}
    \centering
    \includegraphics[width=1.\linewidth]{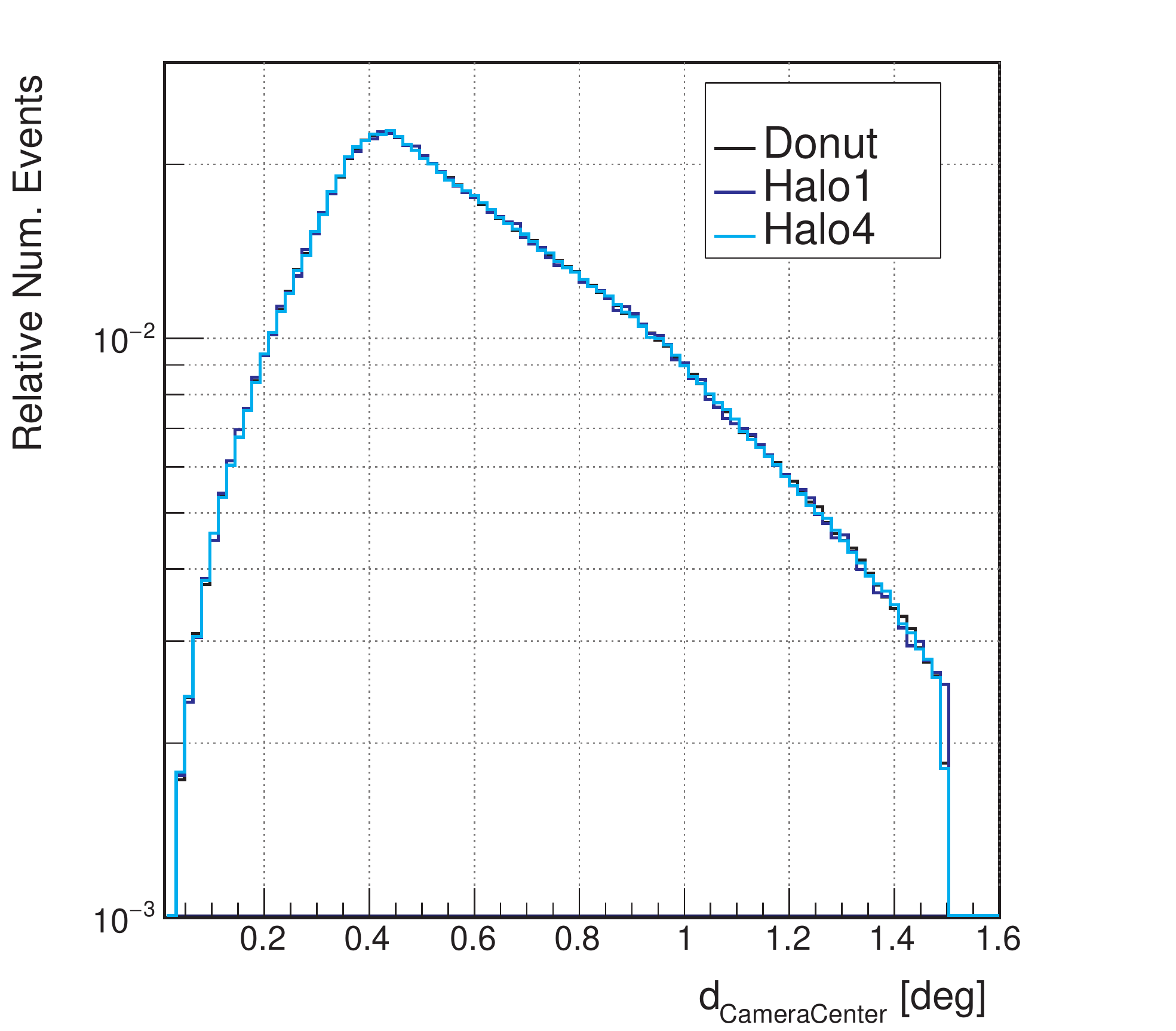}
    \label{fig:DonutMonteCarloCheckOffset}
  \end{subfigure}%
  \begin{subfigure}{.5\textwidth}
    \centering
    \includegraphics[width=1.\linewidth]{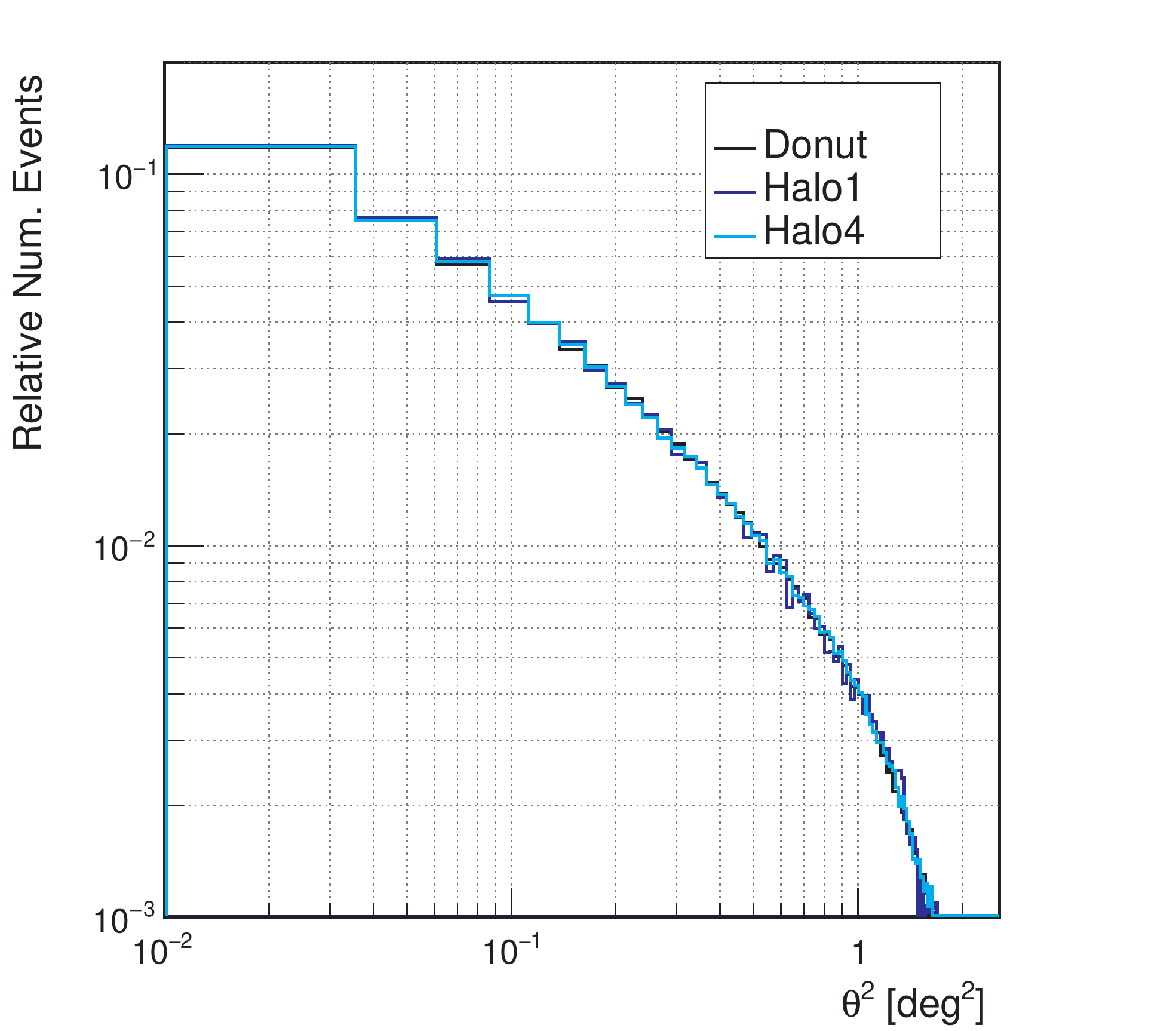}
    \label{fig:DonutMonteCarloCheckTheta2}
  \end{subfigure}
  \end{center}
  \caption{Comparison
    between halo1 (red), halo4 (pink) and donut MC (green). The left plot
    shows the distribution of events as a function of the distance to the
    pointing direction. The right plot shows the distribution of distances
    squared to the associated source center.}
  \label{fig:HaloCrossCheckOffset}
\end{figure}

\begin{figure}
  \centering
  \includegraphics[width=.5\linewidth]{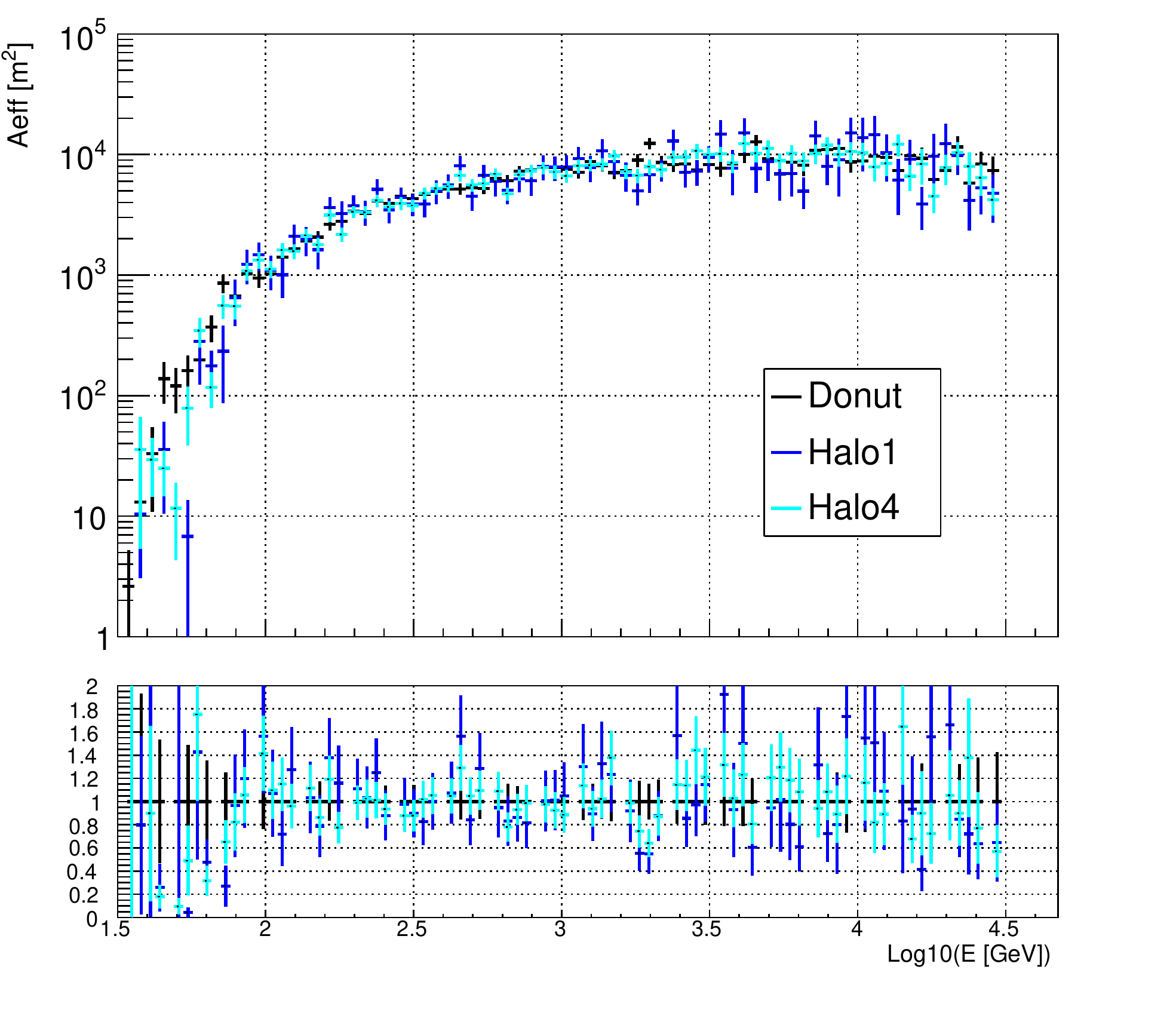}
  \caption{Comparison of Aeff (top) and Aeff ratio with respect to the donut MC (bottom) vs. Etrue computed from halo1, halo4 and donut MC.}
  \label{fig:effectiveAreaComparisonDonutvsHalo}
\end{figure}

\begin{figure}
  \begin{subfigure}{.5\textwidth}
    \centering
    \includegraphics[width=1.\linewidth]{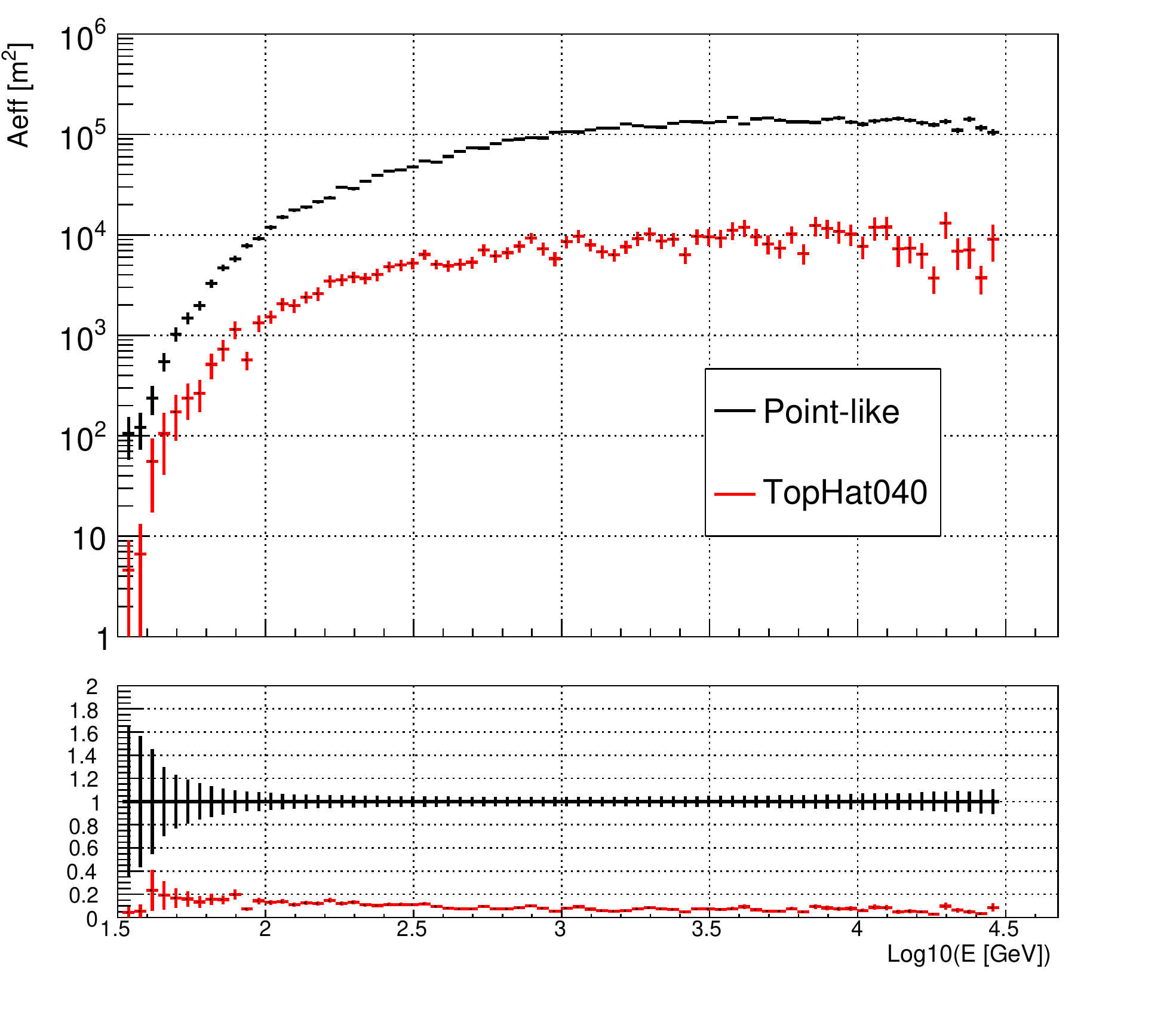}
    \label{fig:effectiveAreaComparisonDelta040vsRing}
  \end{subfigure}%
  \begin{subfigure}{.5\textwidth}
    \centering
    \includegraphics[width=1.\linewidth]{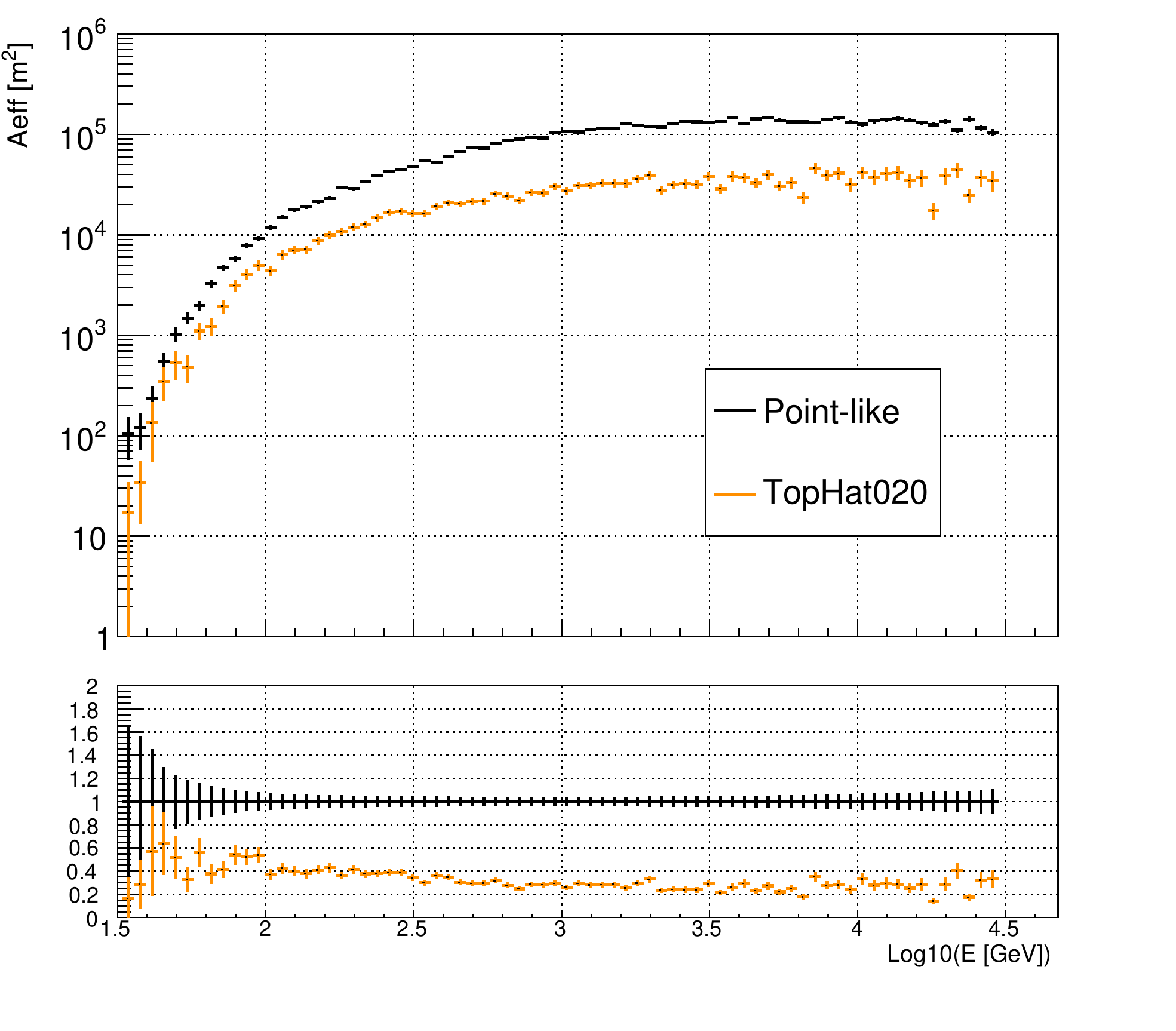}
    \label{fig:effectiveAreaComparisonDelta020vsRing}
  \end{subfigure}
  \begin{subfigure}{.5\textwidth}
    \centering
    \includegraphics[width=1.\linewidth]{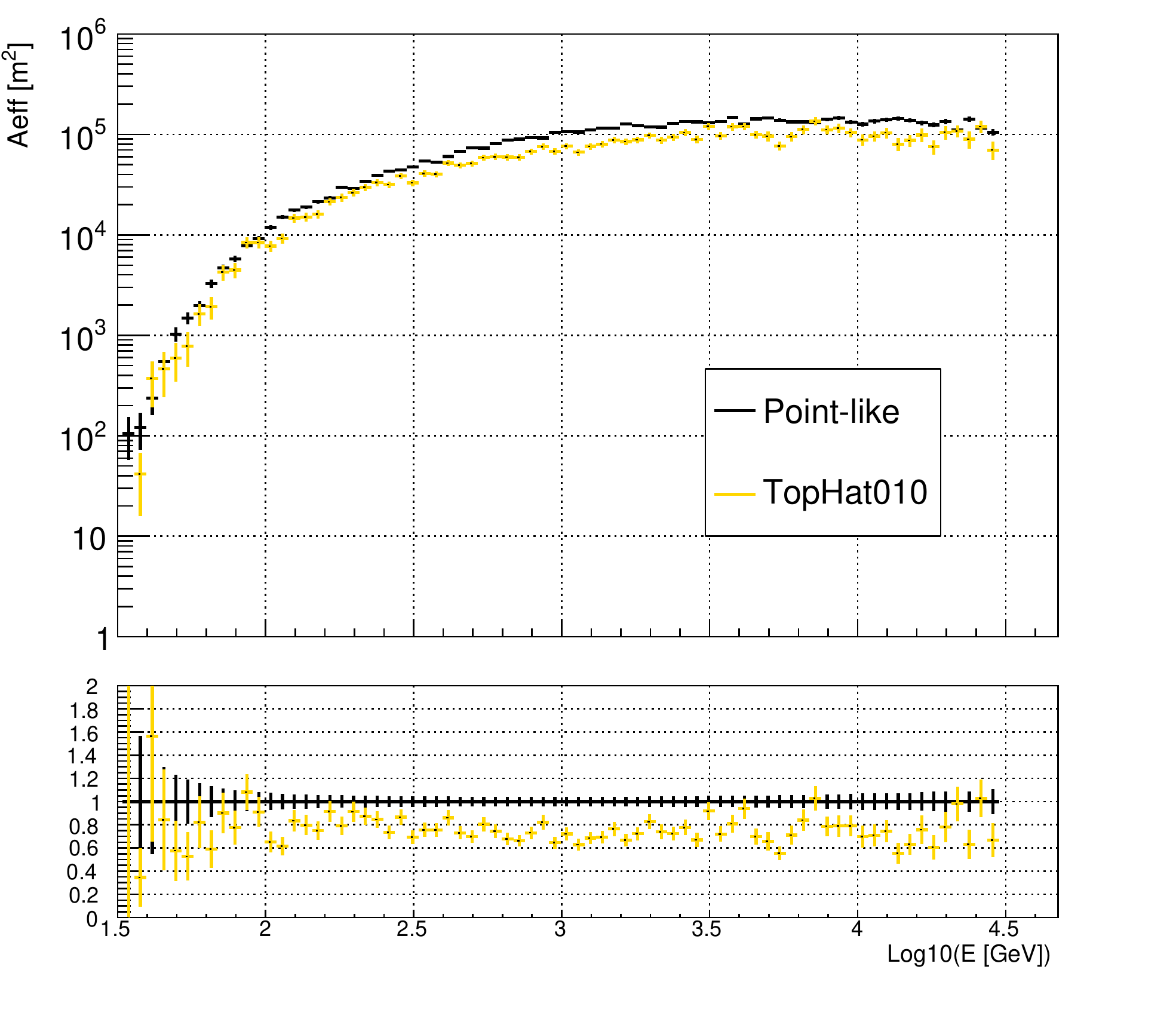}
    \label{fig:effectiveAreaComparisonDelta010vsRing}
  \end{subfigure}%
  \begin{subfigure}{.5\textwidth}
    \centering
    \includegraphics[width=1.\linewidth]{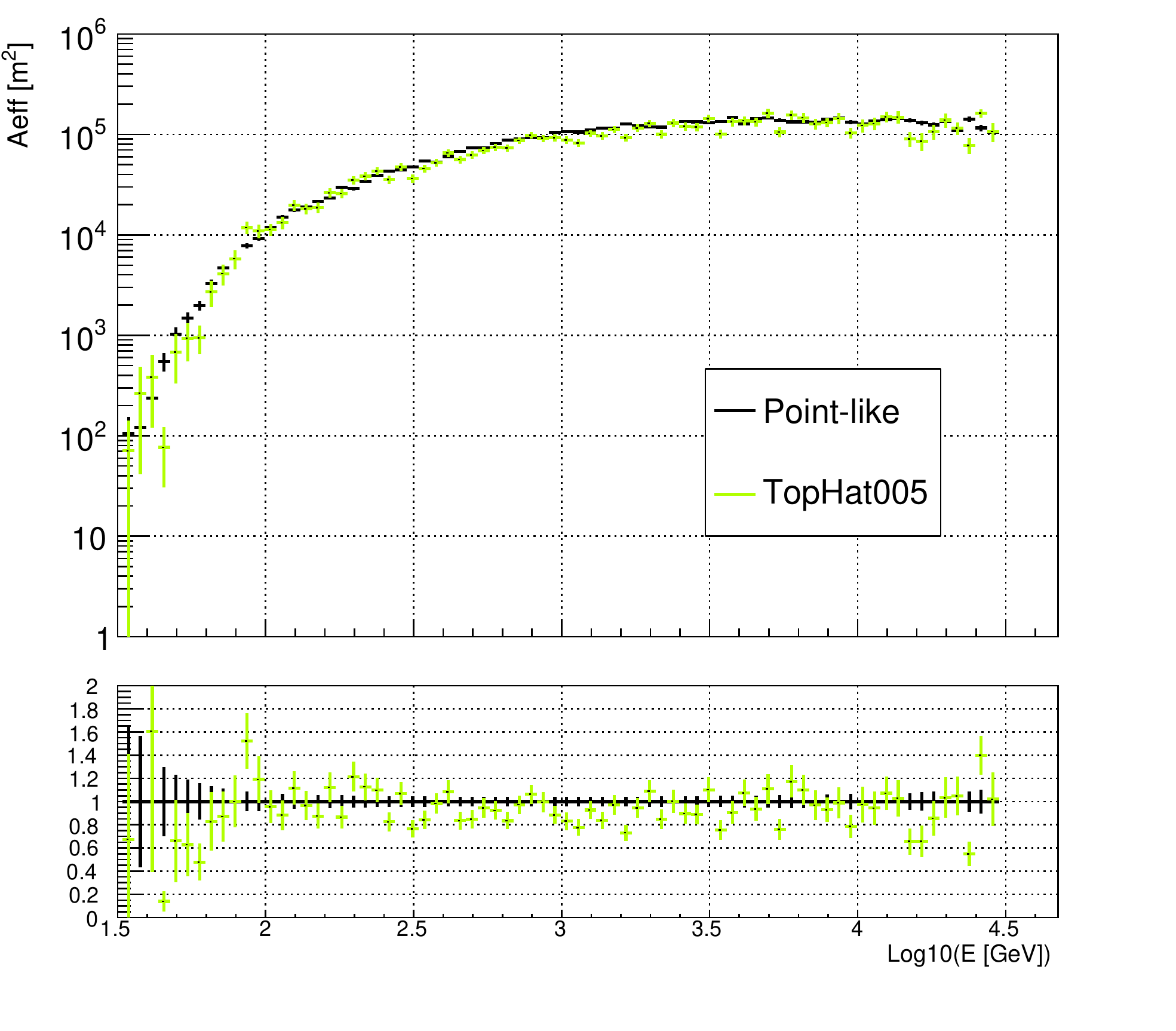}
    \label{fig:effectiveAreaComparisonDelta005vsRing}
  \end{subfigure}
  \caption{Comparison of Aeff (sub panel top) and Aeff ratio with respect to the point-like MC (sub panel bottom) vs. Etrue computed from donut realizations
    using top-hat profiles of 0.4, 0.2, 0.1 and 0.05 degrees (from top bottom
    and left to right, respectively) and point-like MC.}
  \label{fig:DeltaComparison}
\end{figure}

In order to understand the procedure, let us first consider a
simplified version, here named \emph{halo1}, where we select events
from the diffuse MC based on the source morphology for one single,
fixed, orientation between the pointing direction and the source center
(see \autoref{fig:DonutMonteCarloPerseusDecay}, left). 
If we used the \emph{halo1} sample to compute the
IRFs corresponding to the assumed source morphology, we would get the
correct result, but with large statistical uncertainties, given the
relatively low statistics of the selected sample with respect to the
original one, and only valid for one possible orientation between
pointing direction and source center. We can generalize the
\emph{halo1} selection procedure for $n$ halos (halo4, halo10 and
halo100 cases are shown in the right-most plots of 
\autoref{fig:DonutMonteCarloPerseusDecay}). 
If halo-n were constructed simply by repeating the selection procedure of
halo1, the probability of having an event selected more than once,
will get larger, the larger the value of $n$ becomes. In the donut method, this problem 
is solved by selecting diffuse MC events according to a joint probability density function from the
convolution of all possible source center/pointing direction orientations
(see \autoref{fig:DonutMonteCarloPerseusDecay}, most right).
Selected events are associated with a source center randomly chosen
from the expected 0.4$^{\circ}$ ring such that,
at the end of the selection process, all events with a common source center are
spatiallly distributed according to the source morphology.
To show that this procedure works as expected we have performed the
following tests:

\begin{itemize}

\item
Check that the distributions of event directions with respect to the
pointing direction (see \autoref{fig:HaloCrossCheckOffset} left) and the 
associated source center (see \autoref{fig:HaloCrossCheckOffset} right) agree, 
within statistical uncertainy, for the halo-n
and donut realizations. The halo-n and donut distributions show very
good agreement, and a reduction of statistical uncertainty with
growing $n$.

\item
Check that the effective area as a function of the true gamma-ray
energy agree, within statistical uncertainties, computed
for halo-n and donut realizations (see \autoref{fig:effectiveAreaComparisonDonutvsHalo}).
The halo-n and donut distributions
show very good agreement, and a reduction of statistical uncertainty
with growing $n$.


\item
We also expect the IRFs computed with the donut method to converge to
those for a point like MC when we use a very narrow source morphology.
In order to check this, we have produced four different donut
realizations taking the expected distribution true 
directions morphology to be a top-hat function
with radius 0.4, 0.2, 0.1 and 0.05 degrees, respectively, placed at a
wobble distance of 0.4$^{\circ}$.
\autoref{fig:DeltaComparison} shows the comparison of Aeff vs.
Etrue between these four realizations compared to the Aeff vs. Etrue
obtained from the point-like MC. Differences are smaller for smaller
values of the radius, with almost perfect convergence between the 0.05$^{\circ}$
radius halo and the point-like MC.

\end{itemize}

\end{document}